\documentclass[aps,10pt,a4paper,showpacs,pre,twocolumn,            superscriptaddress]{revtex4-1} 

\usepackage{chemformula} 
\usepackage[T1]{fontenc} 
\usepackage{amssymb}
\usepackage{mathrsfs}
\usepackage{stmaryrd}
\usepackage{amsmath}
\usepackage{subfigure}

\usepackage{graphicx}
\usepackage{color}

\usepackage{enumerate}
\usepackage{fancyhdr}
\usepackage{multirow}
\usepackage{hyperref}

\usepackage{xr}
\newcommand{\mr}{\mathrm}

\definecolor{amber}{rgb}{1.0, 0.49, 0.0}

\newcommand{\fs}[1]{{\color{black}#1}}
\newcommand{\fsdel}[1]{}

\newcommand{\xcc}[1]{\textcolor{black}{#1}}

\newcommand{\xc}[1]{\textcolor{black}{#1}}

\newcommand{\rem}[1]{}

\renewcommand{\thesection}{\Roman{section}}

\makeatletter
\renewcommand{\p@subsection}{}
\makeatother

\begin{document}

\author{Xinxiang Chen}
\email{xichen@uni-mainz.de}
\affiliation{Institute of Physics, Johannes Gutenberg-University, 55099 Mainz, Germany}

\author{Lennart Hebestreit}
\affiliation{Institute of Physics, Johannes Gutenberg-University, 55099 Mainz, Germany}



\author{Friederike Schmid}
\email{friederike.schmid@uni-mainz.de}
\affiliation{Institute of Physics, Johannes Gutenberg-University, 55099 Mainz, Germany}

\title[An \textsf{achemso} demo]
  {
  From real polymers to random graphs: percolation thresholds in associative polymer solutions
}





\begin{abstract}

Sol-gel transitions are ubiquitous in soft matter and biological systems, yet their thresholds are often poorly captured by classical Flory-Stockmayer theory because spatial organization and loop formation are neglected. Here, we combine molecular dynamics simulations with random graph and random geometric graph models to determine the respective roles of topology and geometry in reversible associative polymer solutions. We show that a coordinate-free random graph recovers the mean-field Flory-Stockmayer limit, whereas a random geometric graph quantitatively reproduces the shifted percolation thresholds observed in \fs{molecular dynamics} simulations when the detection radius is chosen according to the polymer conformational size.\fsdel{We further demonstrate that this} \fs{This} geometric mapping remains quantitatively valid for linear chains with regularly spaced binding sites over a broad range of chain stiffness. At the microscopic level, we identify primary loops formed already in the pre-gel regime as the dominant \fsdel{origin} \fs{source} of the deviation from mean-field predictions.  Near the gel point, the cluster-size statistics obtained from simulations and random geometric graphs are consistent with the universality class of three-dimensional percolation. These results establish random geometric graphs as a minimal predictive framework for \fs{describing topological transitions in} reversible associative polymer solutions and show that gelation and network formation can be inferred directly from single-chain conformational information.

\end{abstract}

\maketitle

\section{Introduction}

Associative polymers form transient networks through reversible bonding between binding sites distributed along polymer backbones\cite{pappu2023phase}. Depending on concentration\cite{formanek2021gel,semenov1998thermoreversible}, valence\cite{choi2020generalized,rovigatti2022designing}, and bond strength\cite{chen2025sol}, these systems can undergo a sol--gel transition marked by the emergence of a system-spanning cluster. Such percolated networks play a central role in polymer gels, supramolecular materials, and biomolecular condensates, where connectivity governs macroscopic properties such as elasticity, transport, and relaxation\cite{rapp2018mechanisms,rossow2014relaxation,dar2024biomolecular}. Classical theories of gelation, most notably the Flory--Stockmayer (F--S) framework\cite{Flory1941,stockmayer1943theory,stockmayer1944theory}, provide the foundational description of how functionality and reaction probability control network formation in the systems. Within mean-field \fsdel{picture}\fs{theory}, gelation occurs when the average number of new branches generated \fsdel{after arriving at}\fs{upon reaching} a junction becomes at least one, and finite-cluster statistics can be derived analytically under the assumption of tree-like connectivity.

Despite its conceptual importance and widespread use, the classical F--S theory often provides inaccurate quantitative results for realistic associative polymer solutions\cite{mann2019universal,beech2023kinetics}. Real polymer networks are embedded in three-dimensional space: bond formation requires physical proximity, and the connectivity is strongly influenced by conformational correlations and loop formation\cite{beech2023kinetics,wang2017kinetic}. In particular, cycles consume reactive groups without creating new connections that contribute to cluster expansion. As a result, the gel point is shifted relative to the mean-field prediction. Although several corrections based on effective functionality or intramolecular cyclization have been proposed\cite{dobrynin2004phase,santra2021universal,wang2017kinetic}, a simple framework that can capture both the phase behavior and the network-structural properties of complex associative polymer systems is still highly desirable.

Graph-based descriptions provide a natural language for addressing this problem\cite{kryven2016random,zhang2026computational,li2024structural}. Treating each polymer chain as a vertex and each reversible crosslink as an edge maps gelation onto a percolation transition on a graph. In this representation, the classical F--S theory is closely related to a random branching process and can be reproduced by random graph (RG) model in which all vertices are equally accessible and each independent bonding is spatially unconstrained\cite{erdos1959publicationes}. However, this idealized picture neglects a key physical ingredient of polymer solutions: two chains can only bind if they are sufficiently close in space. This observation motivates the use of random geometric graph (RGG)\cite{DallChristensen2002}, in which connectivity is restricted by a finite detection radius. Such a construction offers a minimal way to incorporate spatial locality while preserving the graph-theoretic description of network growth.

In this work, we combine coarse-grained molecular dynamics (MD) simulations with \fsdel{random graph} \fs{RG} and \fsdel{random geometric graph} \fs{RGG} models to investigate percolation in both homo- and hetero-associative polymer solutions. This allows us to compare the roles of topology, spatial constraints and loop formation for sol--gel transition within a unified framework. The RG model serves as a topological reference that recovers the classical F--S mean-field limit \fsdel{exactly} \fs{in the thermodynamic limit}, while MD simulations reveal shifted percolation thresholds relative to this loop-free baseline. We \fsdel{further} demonstrate that these shifts are quantitatively captured by an RGG description when the detection radius is chosen on the basis of polymer conformations measured in simulation.

\fsdel{Beyond the threshold, we} \fs{We further} identify primary-loop formation as the \fs{dominant} microscopic origin of the deviation from mean-field theory. In the homoassociative system, the structure of primary loops is intrachain binding. In the heteroassociative system, owing to the absence of A--A and B--B bonding in our \fsdel{model} \fs{system}, \fsdel{it} \fs{a primary loop} is a minimal cycle formed by \fsdel{repeated} \fs{two} A--B pairings \fs{between same chains}. Introducing a loop-corrected \fsdel{Flory–-Stockmayer}\fs{F--S} criterion via effective-functionality renormalization can improve the \fsdel{fit of}\fs{quality of the analytical} predictions \fs{compared} to \fsdel{real} simulations.

As chain stiffness increases, loop formation is suppressed, causing \fs{both} the simulation results \fs{and the RGG results} to approach the classical \fsdel{Flory--Stockmayer}\fs{F--S} limit\fsdel{, while the random geometric graph}\fs{. The RGG} description remains quantitatively valid over \fsdel{a broad} \fs{the whole} range of chain rigidities. 

We also \fs{consider cluster-size distributions and the giant-cluster fraction, and show how to} derive analytical \fs{mean-field} expressions \fsdel{for the cluster-size distribution and giant-cluster fraction} \fs{for them} using generating functions and the Lagrange inversion theorem. \fsdel{Analysis} \fs{The analysis} of the \fs{real} cluster-size distribution near the gel point \fsdel{further} reveals a crossover\fsdel{in universality class} \fs{between universality classes}: the \fsdel{random graph} \fs{RG} model retains mean-field critical behavior, whereas molecular dynamics and \fsdel{random geometric graph} \fs{RGG} results are consistent with three-dimensional percolation scaling. 

\fsdel{Taken together, our} \fs{Our} results show that realistic associative polymer solutions can be understood as stochastic branching processes constrained by finite-range \fsdel{geometry} \fs{geometric constraints}, with RGGs providing a minimal predictive bridge \fsdel{between} \fs{connecting} polymer conformation\fsdel{and} \fs{to} gelation behavior.

\section{Model and methods}

\subsection{Molecular dynamics simulation}
\label{subsec:MD_simulation}

In the\fsdel{framework of} molecular dynamics simulations, we model associative polymer\fs{s} in good solvent using \fsdel{the} \fs{a Kremer-Grest type} bead-spring model\cite{grest1986molecular} \fs{in implicit solvent}.\fsdel{The bead} \fs{Beads} along a chain are connected by \fs{a} standard\fsdel{Kremer--Grest potential (KG model)} \fs{nonlinear FENE-potential} \cite{grest1986molecular,kremer1990dynamics}, with spring constant $k_b=30k_B T/\sigma^2$ and $R_0=1.5\sigma$. 
To study the effect of stiffness on the percolation transition, \fsdel{the bond angle potential is also employed in a harmonic manner:} \fs{we additionally introduce an angular potential} $U_\mr{bending}(r)=\frac{1}{2}k_a(\theta-\pi)^2$, where $\theta$ is the angle between consecutive bonds along a chain. All nonbonded interactions between beads\fsdel{(excluding interactions} \fs{except} between \fsdel{different species of} binding sites \fsdel{)} are purely repulsive Weeks-Chandler-Andersen (WCA) \fs{interactions with the pair} potential\cite{weeks1971role}, $U_\mr{WCA}(r)=4\epsilon\left[(\sigma/r)^{12}-(\sigma/r)^6+1/4\right]$ when $r<2^{1/6}\sigma$ and 0 otherwise. \fsdel{except for a short-range specific attractive potential between A--B binding sites that enforces one-to-one binding.}

\fsdel{To mimic the homoassociative and heteroassociative polymer system, two models are utilized to achieve the specific one-to-one binding in the simulation.}
\fs{To implement specific one-to-one binding, we introduced a special type of monomers (`binding sites') that can form specific bonds and are evenly distributed along the chains, separated by linker chains made of 
neutral beads. Two types of interaction models are utilized to implement specific bonding, which are both based on the short-range attractive potential} $U_\mr{b}(r)=-\epsilon_\mr{sp}\left[1+\cos(\pi r/0.5\sigma)\right]$ when $r<0.5\sigma$ and 0 otherwise.

\begin{figure}[ht]
\includegraphics[width=8cm]{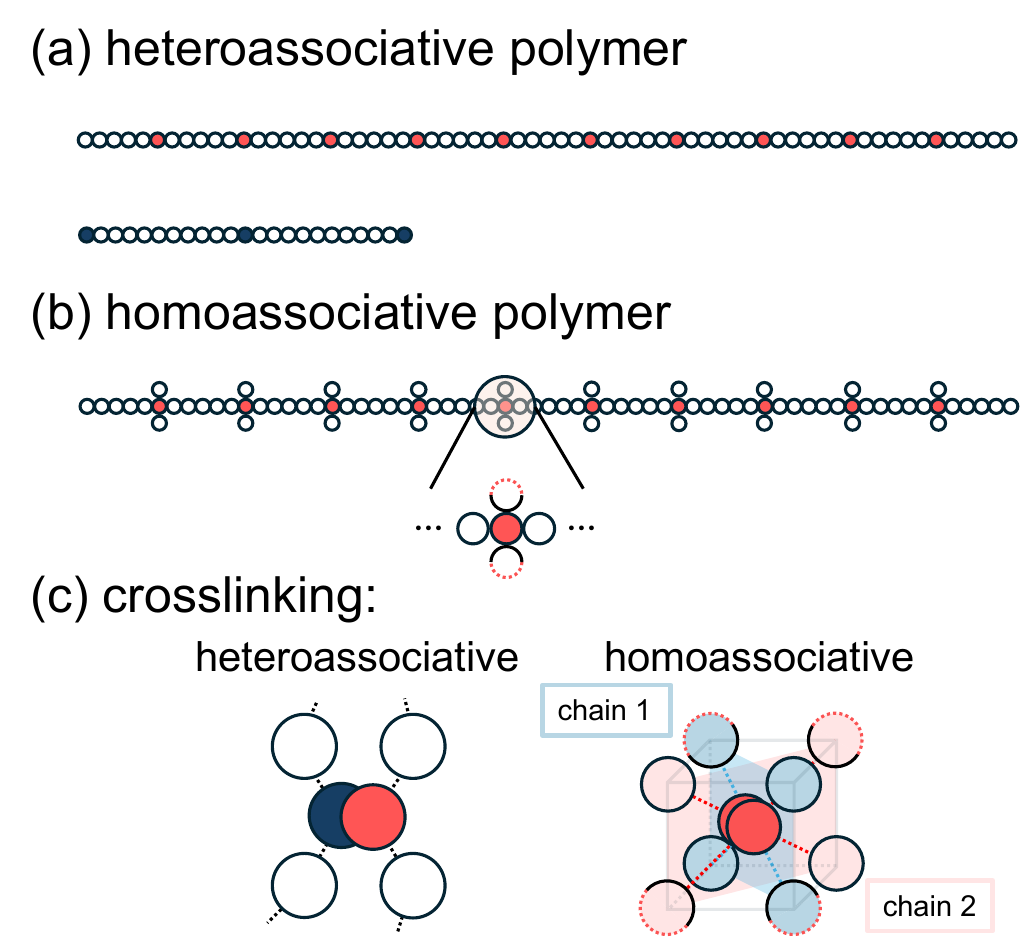}
\caption{The structure of heteroassociative (a) and homoassociative (b) polymers in the \fsdel{paper}\fs{simulation model}. The red and blue beads are the binding sites for each component, connected by the neutral linkers (white beads). For the homoassociative polymer in (b), each binding site contains two side beads, which \fs{ensure one-to-one binding by their steric interactions}\fsdel{have a steric effect to achieve one-to-one binding}. Side beads \fsdel{are neutral type, while they are ghost particles for binding sites(carry no interaction with any binding sites)}\fs{have repulsive interactions with all other beads except the central beads of binding sites}. (c) \fsdel{The details of crosslinkings for both associative polymer systems} \fs{Cartoons showing configurations of bound binding sites}.}
\label{fgr:structure}
\end{figure}

For the heteroassociative polymer solution,  \fsdel{the two types of polymers carry distinguishable binding sites A and B, which are evenly distributed along the chain and separated by the neutral linkers.} \fs{the binding sites are single, distinguishable beads A and B. Pairs of A--B beads attract each other with the
pair potential $U_b(r)$, A--A and B--B have longer-range purely repulsive WCA interactions. This ensures one-to-one binding\cite{zhang2021decoding,chen2025sol}.}
\fsdel{The architecture of the A and B chains used in the system are set (shown as (Figure~\ref{fgr:structure}(a)): chain length, the number of binding sites, and the number of the neutral linkers between two binding sites for A is $N_A=65, f_A=10$, $l_A=5$; for B: $N_B=23, f_B=3, l_B=10$.} 

For homoassociative polymers, all binding sites are identical \fs{and we must use a different approach to enforce one-to-one binding}.
\fsdel{Within the previous model, introducing short-range attraction between identical binding sites necessarily removes the mutual repulsion. This change of repulsion allows multiple same-type sites to simultaneously interact within the binding range, thereby preventing one-to-one binding.} A common strategy \fsdel{to enforce one-to-one binding in homoassociative systems }is using a repulsive three-body interaction\cite{sciortino2017three, rovigatti2022designing, rovigatti2023entropy}. \fs{However, in our system, this lead to instabilities at higher densities. Therefore, we instead modeled as monomers consisting of three beads, one central bead, which attracts
other central beads, and two side beads which sterically prevent central beads from participating in more than one bond (see Figure~\ref{fgr:structure}(c)). Side beads
have repulsive WCA interactions with each other and with neutral beads. Center beads
interact via the potential $U_b$. Using this model, the system can}
\fsdel{Here, instead of using explicit three-body interactions, we introduce a special structure of each binding site, as illustrated in Figure~\ref{fgr:structure}(b). In this model, there are two neutral side beads attached to the binding site, which generates a sterically crowded local packing(body-centered cubic-like packaging), when the specific binding forms (shown in Figure~\ref{fgr:structure}(c)). These side beads do not interact with any binding sites. Therefore, we use the half red dashed contour line to indicate that these are ghost beads for the binding sites in Figures~\ref{fgr:structure}(b,c). However, they interact with other neutral beads through the same pure repulsive interactions as ordinary neutral beads. In this way, the side beads introduce steric hindrance without directly altering the pairwise attractive interaction between binding sites. The system can }maintain effective one-to-one binding \fsdel{even }at \fsdel{a }total monomer \fsdel{density}\fs{densities} as high as $c=0.8/\sigma^3$. 
\fsdel{The details of the homoassociative chains in the system are set: chain length, the number of binding sites, and the number of the neutral linkers between two binding sites for A is $N_\mr{homo}=65+2\times 10=85$, $f_\mr{homo}=10$, $l_\mr{homo}=5$.}

\fs{The parameters in the simulations were as follows. 
In the homo-associative system, we considered chains with $f_\mr{homo}=10$ binding sites per chain separated by linkers containing $l_\mr{homo}=5$ beads. The total number of beads per chains was $N_\mr{homo}=65+2\times 10=85$ beads per chain.
In the hetero-associative system, we considered mixtures of $A$ and $B$ chains
of length $N_A = 65$ and $N_B = 23$ with $f_A =10$ and $f_B=3$ binding sites, separated by linkers of length $l_A=5$ and $l_B = 10$, respectively.}

All simulations are performed using HOOMD-blue (v2.9.3 and v3.7.0)\cite{anderson2008general} in a cubic box of volume $(80\sigma)^3$ \fs{with periodic boundary conditions}. All monomers have the same mass and diameter $\sigma$, and the WCA energy scale is fixed at $\epsilon=1k_B T$. For the heteroassociative system, the specific binding strength is fixed at $\epsilon_\mr{sp}=6k_B T$, which ensures frequent bond opening and closing consistent with reversible crosslinking. For the homoassociative system, the binding strength is varied over the range $\epsilon_\mr{sp}=3$--$26k_B T$. The larger value required in the homoassociative case compensates for the steric penalty introduced by the side beads, which effectively weakens \fs{the} binding strength. Each system was equilibrated \fsdel{for $10^6t_0$}\fs{over a time $0.8 \times 10^6t_0$ in simulation units $t_0 = \sqrt{m \sigma^2/k_B T}$, and data were collected over $0.2 \times 10^6t_0 $.} 

\subsection{Random graph and Random geometric graph}
\label{subsec:graph_theory}

We construct both RG and RGG models for the associative polymer systems studied here. In these graph representations, each polymer chain corresponds to a vertex, and the number of binding sites(functionality) on each chain defines the maximum allowed degree of the vertex.

\begin{figure*}[ht]
\includegraphics[width=13cm]{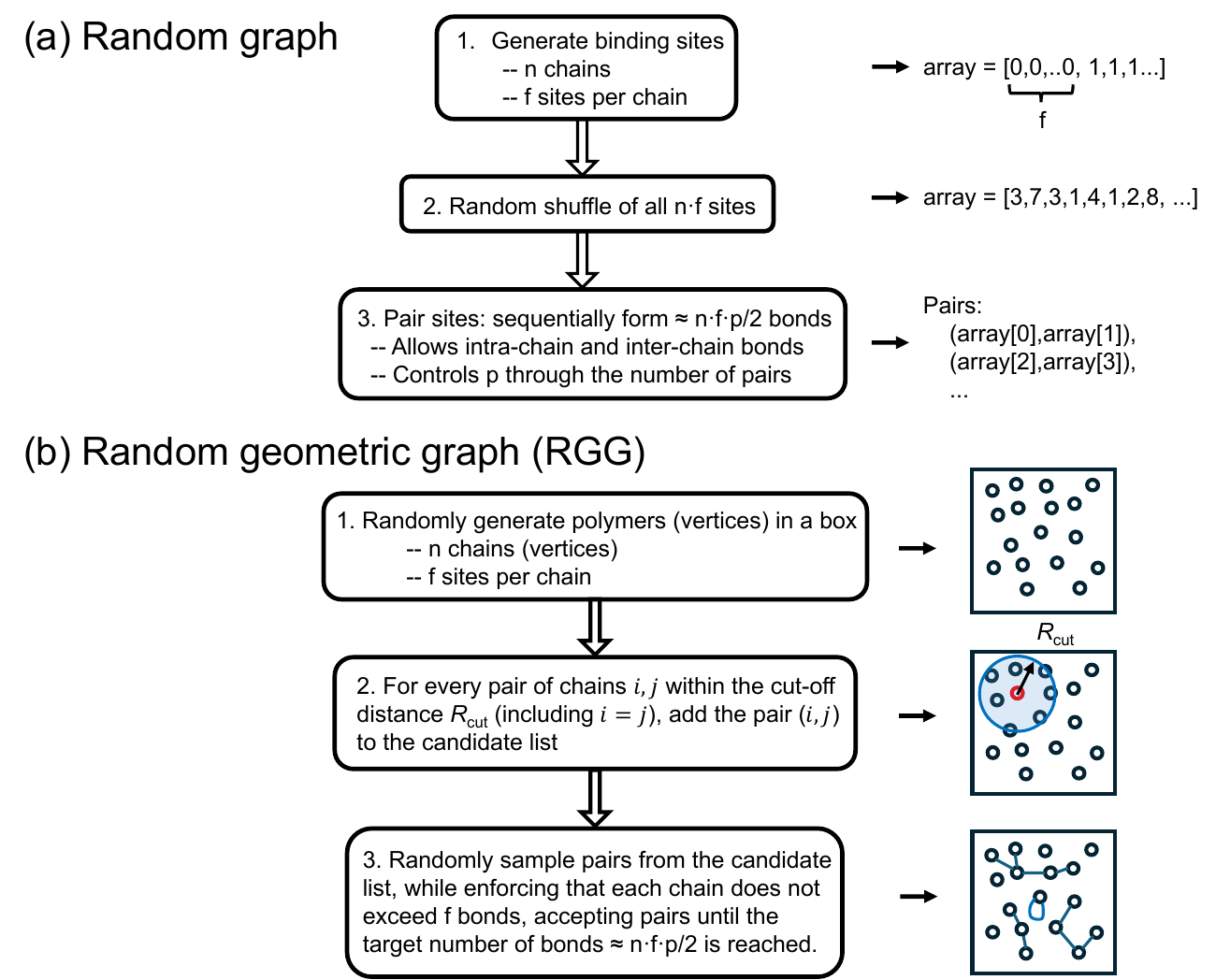}
\caption{Flowchart of the random graph(RG) (a) and random geometric graph (RGG) \fs{generation} (b) for the homoassociative polymer system. The procedure for heteroassociative polymers \fs{(not shown)} is analogous, except that a second type of vertices is introduced, and the bond formation is restricted to occur only between two different kinds of vertices (i.e., only A--B crosslinking is allowed, while A--A and B--B crosslinking are forbidden). \fsdel{For brevity, the corresponding flowchart for the heteroassociative case is not shown.}
}
\label{fgr:flowchart}
\end{figure*}

We first consider the RG model, which generalizes the classical Erd\H{o}s--R\'enyi framework\cite{erdos1959publicationes}. In this model, bond formation is purely \fsdel{topological:}\fs{random:} spatial constraints are neglected, and all vertices are equally accessible for bonding. For a homoassociative system of $n$ identical chains with functionality $f$, the network is represented by a graph $G(n,f,p)$, in which a total of $nf$ binding sites are randomly paired with binding probability $p$. Algorithmically, all binding sites are first enumerated and randomly scrambled, and pairs are then sequentially connected until the total number of bonds reaches $nfp/2$ (see Figure~\ref{fgr:flowchart}(a)). For heteroassociative systems, the construction is extended to a bipartite random graph $G(n_A,n_B,f_A,f_B,p_A,p_B)$, in which only vertices of type A and B can \fsdel{only} connect. In this case, the stoichiometric constraint
\begin{equation}
p_A f_A n_A = p_B f_B n_B
\end{equation}
ensures consistency between the numbers of reacted binding sites on the two components.

We systematically scan the binding probability $p$ for homoassociative systems and $(p_A,p_B)$ for heteroassociative systems to construct phase diagrams and locate the percolation threshold. Because the RG model has no spatial embedding, a spanning-cluster criterion cannot be applied directly. We therefore adopt a graph-based operational criterion\fsdel{ in RG model}, defining the transition as the point at which the largest connected component exceeds $3\%$ of all vertices. In practice, this threshold corresponds to the point at which the largest cluster begins to grow rapidly. This is analogous to identifying a finite-size transition from the \fsdel{emergence}\fs{onset of rapid growth} of an order parameter in models such as the Ising model. Since the RG model neglects spatial correlations, it serves as a purely topological reference corresponding to the mean-field limit of network formation.

To incorporate the effect of spatial locality, we further construct \fsdel{random geometric graph}\fs{RGGs}\cite{dall2002random}. In this model, \fs{we assign positions in three-dimensional space to vertices, and} bond formation is restricted by finite-range geometric accessibility: two polymers can only bind if they are sufficiently close in space. Algorithmically, we first place $n$ polymers (vertices) randomly in a three-dimensional cubic box. We identify all possible vertex pairs $(i,j)$ whose center-to-center distance satisfies $r_{ij}\le R_\mr{cut}$, where $R_\mr{cut}$ is the detection radius, and add them to a candidate list (see Figure~\ref{fgr:flowchart}(b)). The $r_{ij}$ \fsdel{is}\fs{are} calculated using the minimum-image convention under periodic boundary conditions, so this model is consistent with the standard definition of random geometric graphs in a continuous periodic domain\cite{dall2002random}. \fsdel{For}\fs{In} the homoassociative system, the case $i=j$ is also included in order to allow intrachain pairing events. Candidate pairs are then sampled randomly, and a bond is accepted only if neither polymer has exceeded its maximum number of allowed bonds (functionality). The acceptance procedure is repeated until the target number of bonds, approximately $nfp/2$, is reached. In this way, the RGG retains the same functionality constraint as the RG model while additionally imposing a finite interaction range.

Because the RGG depends explicitly on particle positions, \fs{the} percolation transition is identified through the formation of a system-spanning cluster under periodic boundary conditions\cite{chen2025sol}. The definition of the spanning cluster is given in Section~\ref{subsec:threshold}. In the limit $R_\mr{cut}\rightarrow L$, where $L$ is the box size and all vertex pairs become geometrically accessible, the RGG reduces to the corresponding RG model.

For both RGs and RGGs, we identify connected components of the given graph, determine the cluster-size distribution \fsdel{$n(s)$}\fs{$n_s$}, and calculate the giant-cluster fraction $G$. \fsdel{Owing to the stochastic nature of the graph construction, all}\fs{All} reported quantities are obtained by averaging over $20$ independent realizations for each parameter set. To quantify cycle formation in the following discussion, \fsdel{we estimate the total number of  cycles in each graph by combining two contributions. First,} we compute the cyclomatic number\cite{newman2018networks}, $N_\mr{cyc} = E - V + C$, where $E$, $V$, and $C$ denote the numbers of edges, vertices, and connected components, respectively. This quantity counts the number of independent cycles in a \fsdel{simple} graph,\fsdel{ i.e., a graph without} \fs{also accounting for} self-loops or multiple edges. \fsdel{Second, we explicitly include primary loops that are not captured by the cyclomatic number, namely self-loops in the homoassociative case and multiple edges between the same pair of vertices in the heteroassociative case (see Figure~\ref{fgr:loop_all}(a)). These local cyclic motifs are excluded from the definition of a simple graph and must therefore be counted separately and added back to the cyclomatic number to calculate the total cycle number.}

\subsection{Theory}
\label{subsec:theory}

The sol--gel transition can be estimated using analytical approaches derived from the Flory--Stockmayer theory, which provides a foundational description of network formation and gelation in \fsdel{associating}\fs{associative} polymer systems\cite{flory1953principles,stockmayer1943theory,semenov1998thermoreversible,danielsen2023phase}. In the present work, we consider both homoassociative and heteroassociative polymer networks. For \fsdel{the} systems with fixed functionality of each \fsdel{associative} polymer, the corresponding threshold conditions are
\begin{equation}
p_{\mathrm{c}} = \frac{1}{f-1},
\label{homo_pc}
\end{equation}
for the homoassociative case, and
\begin{equation}
m=p_Ap_B(f_A-1)(f_B-1)=1,
\label{hetero_pc}
\end{equation}
for the heteroassociative case. These expressions rely on the classical tree-like assumption and therefore neglect \fsdel{loop}\fs{cycle} formation.

\fsdel{Equivalent}\fs{The same} threshold conditions can also be derived using probability--generating functions for stochastic branching processes, as shown by Gordon et al.\cite{gordon1962good}. In this approach, one considers the probability that a branch reached by following a bond terminates after a finite number of steps. In the sol state, all connected clusters are finite, so this probability is unity. Gelation occurs when an infinite cluster first appears, causing this probability to deviate from unity.\fsdel{This approach yields the same threshold condition as the F--S theory, as shown in Appendix~\ref{derive_GF}.} \fs{We apply this approach to the polymer systems considered  here in Appendix~\ref{derive_GF},
and show that it yields the same threshold conditions as the F--S theory.}

\fsdel{Thus, besides}\fs{Apart from} the F--S and generating--function descriptions, gelation can \fs{simply} be identified from the divergence of the weight-averaged molecular weight $M_w$ \fs{of the largest cluster}. This formulation is not only more direct, but  \fsdel{also more convenient for extension}\fs{can also be extended more easily} to systems with \fsdel{functionality polydispersity}\fs{polydisperse functionalities}. 
Using the law of conditional expectation, Macosko and Miller showed that the gel point depends on the average functionality of the network-forming species\cite{macosko1976new}.  \fsdel{Following the framework of Macosko and Miller, the} \fs{The} generalised  \fs{percolation} threshold for homoassociative systems \fsdel{becomes}\fs{is given by}
\begin{equation}
p_\mr{c} = \frac{1}{\bar f-1},
\label{homo_pc_generalized}
\end{equation}
where $\bar f=\sum_i x_i f_i$ is the average functionality of the polydisperse system, with $x_i$ denoting the fraction of components with functionality $f_i$. For heteroassociative systems with selective binding between different components, the corresponding generalized condition is
\begin{equation}
m = p_Ap_B(\bar f_{A}-1)(\bar f_{B}-1)=1,
\label{hetero_pc_generalized}
\end{equation}
where $\bar f_A$ and $\bar f_B$ are the average functionalities of components A and B, respectively. The detailed derivation is given in Appendix~\ref{derive_pc_generalized}.

\fsdel{Motivated by the generalized threshold conditions derived above, 
we}\fs{The generalized threshold conditions can be used to estimate the effect of 
loop formation on the percolation threshold. Recently}
\fsdel{More recently}, kinetic Monte Carlo simulations \fsdel{also demonstrated}\fs{showed} that primary loops effectively change the junction functionalities, which in turn suppresses gelation\cite{wang2017kinetic}. As will be shown in Section~\ref{subsec:loop}, primary loops are \fsdel{already} present in appreciable amounts \fs{in our systems}, especially in the pre-gel regime. \fsdel{Two corrections describe distinct consequences of loop formation.}\fs{Loop formation has two distinct consequences:} First, a loop consumes \fsdel{part of}\fs{binding sites, which reduces} the molecular functionality. Second, bonds belonging to loops do not \fsdel{contribute fully to the number of}\fs{create new} intermolecular connections. 
\fs{We therefore propose to} incorporate loop effects by \fsdel{correcting}\fs{renormalizing} both the functionality distribution and the binding probability, \fs{and}\fsdel{We therefore} define a loop-corrected functionality for each molecule and a corresponding loop-free binding probability.

For the homoassociative system shown in Figure~\ref{fgr:loop_all}(a), a primary loop is an intrachain bond and consumes two binding sites on the same chain. The average effective functionality is
\begin{equation}
\left\langle f_{\mr{eff}}\right\rangle=f-\frac{2n_{\mr{loop}}}{n_{\mr{chain}}},
\label{eq:homo_feff_average}
\end{equation}
where $n_{\mr{loop}}$ is the total number of primary
loops, and $n_{\mr{chain}}$ is the total number of chains. The loop-free binding probability is \fsdel{therefore }defined in relation to the remaining bound sites.:
\begin{equation}
p_{\mr{eff}}=\frac{2\left(n_{\mr{bound}}^{\mr{total}}-n_{\mr{loop}}\right)}{n_{\mr{chain}}\left\langle f_{\mr{eff}}\right\rangle}.
\label{eq:homo_peff}
\end{equation}
Together, we can modify gelation threshold to become
\begin{equation}
p_{\mr{eff}}(\left\langle f_{\mr{eff}}\right\rangle-1)=1.
\label{eq:homo_loop_corrected_threshold}
\end{equation}
This expression can be regarded as a specific averaged form of the generalized percolation condition. 
For homoassociative systems, a similar correction for the functionality was previously \fsdel{adopted}\fs{proposed} by Dobrynin\cite{dobrynin2004phase}.

For the heteroassociative system, intrachain binding is absent. The \fs{lowest order cycles, which we call} primary loops, are instead generated by multiple bonds between the same pair of $A$ and $B$ molecules, as illustrated in Figure~\ref{fgr:loop_all}(b). Although a two-edge primary loop contains two $A$--$B$ bonds, one of these bonds still establishes a connection between the two molecules. Only the additional bond is redundant for network branching. More generally, if two molecules are
connected by $\ell_i$ multiple bonds, only one bond establishes an independent molecular connection and the remaining $\ell_i-1$ bonds are redundant.

\fsdel{So}\fs{Thus} the total number of redundant bonds is
\begin{equation}
n_{\mr{red}}=\sum_i(\ell_i-1).
\label{eq:n_redundant}
\end{equation}
The effective functionalities for each component are
\begin{equation}
\left\langle f_{A,\mr{eff}}\right\rangle=f_A-\frac{n_{\mr{red}}}{n_A},
\,
\left\langle f_{B,\mr{eff}}\right\rangle
=f_B-\frac{n_{\mr{red}}}{n_B}.
\label{eq:hetero_feff_average}
\end{equation}
\fsdel{Meanwhile, the}\fs{The corresponding} loop-free binding probabilities for the two components are
\begin{equation}
p_{A,\mr{eff}}=\frac{n_{\mr{bound}}^{\mr{total}}-n_{\mr{red}}}{n_A \left\langle f_{A,\mr{eff}}\right\rangle},\,
p_{B,\mr{eff}}=\frac{n_{\mr{bound}}^{\mr{total}}-n_{\mr{red}}
}{n_B \left\langle f_{B,\mr{eff}}\right\rangle}.
\label{eq:hetero_pB_eff}
\end{equation}
The modified heteroassociative gelation criterion is therefore
\begin{equation}
m^\ast
=p_{A,\mr{eff}} \: p_{B,\mr{eff}} \:
\left(\left\langle f_{A,\mr{eff}}\right\rangle-1\right) \:
\left(\left\langle f_{B,\mr{eff}}\right\rangle-1\right)
=1.
\label{eq:hetero_pc_modify}
\end{equation}
In the absence of loops, $n_{\mr{loop}}=n_{\mr{red}}=0$, the effective functionalities and binding probabilities reduce to their uncorrected values, and the classical Flory--Stockmayer criteria are recovered.

Beyond locating the percolation threshold, we also characterize the network structure by computing the cluster size distribution \fsdel{$n(s)$}\fs{$n_s$} and the fraction of the giant (gel) component $G$. Classical results for \fsdel{$n(s)$}\fs{$n_s$} in branched-polymer gelation can be traced back to Stockmayer's seminal treatment of molecular size distributions in branched polymers\cite{stockmayer1952molecular}. Here we derive \fsdel{$n(s)$}\fs{$n_s$} and $G$ within a unified formalism based on probability generating functions\cite{gordon1962good,newman2001random}, and obtain explicit expressions via the Lagrange inversion theorem\cite{good1960generalizations,good1965generalization,gessel1987combinatorial,flajolet2009analytic,surya2023lagrange}. The details of the derivations are provided in Appendix~\ref{derive_GF}.

\section{Results and discussion}
\label{sec:results}

\subsection{Percolation threshold}
\label{subsec:threshold}

We begin with investigating the phase diagram of the percolation transition for homoassociative and heteroassociative polymer solutions. In the literature\cite{harmon2017intrinsically,danielsen2023phase,chen2025sol},\fsdel{ the} system\fs{s} of associative polymers can exhibit two kinds of phase transition: phase separation and percolation. Phase separation can be suppressed under certain conditions,\fsdel{ such as} \fs{e.g.,} when the binding strength is weak, when the number of binding sites is small \fs{or when excluded volume interactions are strong}. \fsdel{In such systems, percolation transitions occur solely. Here, we choose the proper chain structure of homoassociative and heteroassociative polymer present in Section \ref{subsec:MD_simulation} to independently discuss the percolation behavior in these systems.}\fs{This is the case in the systems considered in our simulations, see Section \ref{subsec:MD_simulation}: they do not phase separate, they only feature percolation transitions.}

\begin{figure}[ht]
\includegraphics[width=8cm]{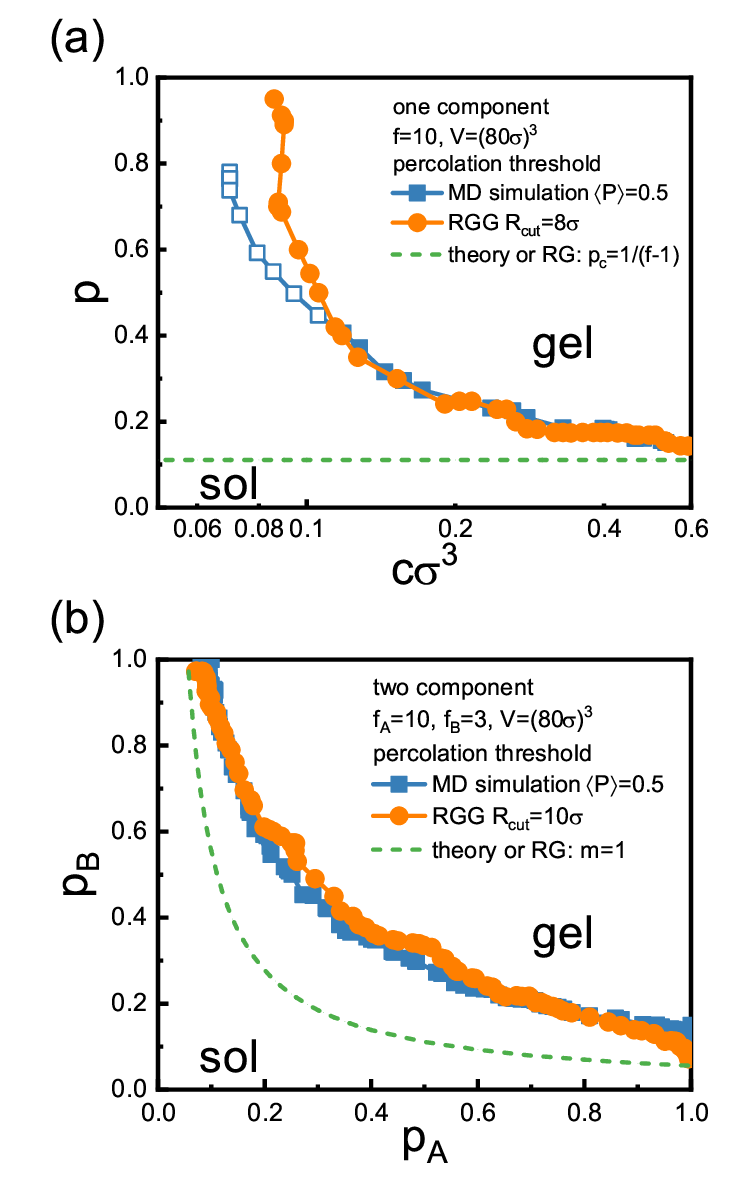}
\caption{Percolation threshold for (a) homoassociative and (b) heteroassociative polymer solution\fs{s}.\fsdel{The threshold line separates the phase diagram into two regions: the} \fs{The} lower-left region corresponds to the sol state, \fsdel{while} the upper-right region \fsdel{corresponds} to the gel state. The blue point lines \fsdel{are the simulation results, in which}\fs{mark parameter values where the} averaged percolation probability \fs{is} $\langle P\rangle=0.5$ \fs{in simulations}. \xc{The open symbols in (a) denote points where binding
energies are so large ($\epsilon_\mr{sp}> 13k_BT$) that the system could not equilibrate (irreversible binding case).}
The orange point lines are RGG results with detection radius $R_\mr{cut}=8\sigma$ for \fs{the} homoassociative case and $R_\mr{cut}=10\sigma$ for \fs{the} heteroassociative case. Green dashed lines show the prediction of F--S theory, \fsdel{ used} Eq.~\eqref{homo_pc} and Eq.~\eqref{hetero_pc}, which coincide with the RG results.
}
\label{fgr:P_c_all}
\end{figure}

\fsdel{To describe the percolation transition, we define it as the emergence of an infinite-size network\cite{livraghi2021exact}. Since our simulations use periodic boundary conditions in a finite-size system, a spanning cluster is defined as one for which} 
\fs{In the thermodynamic limit, the percolation transition is defined via the emergence of an infinite-size connected cluster\cite{livraghi2021exact}. In our finite systems, however, we cannot have infinite-size clusters, therefore we instead consider spanning clusters, defined as clusters where} a closed path can be traced from any particle to its periodic image\cite{chen2025sol}.
Based on this definition, we \fsdel{define}\fs{identify} a percolation \fsdel{probability}\fs{order parameter} $P$ as,
\begin{equation}
\label{eq:P_order}
P=\left\{
\begin{array}{rcl}
&1 &: {\text{system contains $\geq 1$ spanning clusters}} \\
&0&: {\text{otherwise}}
\end{array}\right.
\end{equation}
and its time-averaged value, denoted by $\langle P \rangle$, in equilibrated systems. Figure~\ref{fgr:P_c_all} shows the behavior of percolation in the homo- and hetero-associative systems. For both cases, the percolation threshold is extracted from $\langle P\rangle=0.5$ (blue point line).

\fsdel{Different from previous studies where the phase diagram is expressed in terms of monomer concentration and binding strength $\epsilon_\mr{sp}$ (or equivalently associative rate $K$)
, binding strength is not an optimal variable due to the different microscopic chain model employed here. However, it controls the equilibrium probability of forming bonds at a given concentration. Therefore, in the following discussion, we adopt the binding probability $p$ in equilibrium as the control parameter.}
\fs{In the literature, phase diagrams are often shown as a function of binding strengths. Here, in order to facilitate the comparison of chain models and the comparison with theory, we choose to instead present them in terms of binding probabilities $p$ at equilibrium.} 
For the homoassociative system (Figure~\ref{fgr:P_c_all}(a)), the phase diagram is presented in the ($c$, $p$) plane, where $c$ is the monomer concentration. For the heteroassociative case (Figure~\ref{fgr:P_c_all}(b)), we fix the binding strength and vary two monomer concentrations. \fsdel{As a result, the}\fs{The} phase diagram is thus represented in the ($p_A$, $p_B$) space. When the concentration or binding probability crosses the threshold, a percolation transition occurs,\fsdel{resulting in the system coexisting with} \fs{beyond which the system contains} a spanning cluster(gel \fsdel{state}\fs{fraction}) \fsdel{and}\fs{which coexists with} some smaller clusters(sol \fsdel{state}\fs{fraction})\cite{ranganathan2020dynamic}. \xcc{The corresponding binding strength and concentration utilized in our simulation can be found in Figure S1 of Supporting Information.}

In the homoassociative case(Figure~\ref{fgr:P_c_all}(a)), mean-field theory predicts that the threshold is constant ($p_c=1/(f-1)=1/9$, as shown in green dashed line), while in the real simulation, the formation of a spanning cluster requires a higher binding probability in the system, especially in \fs{the regime of} dilute concentration\fs{s}. In \fs{the} heteroassociative case (Figure~\ref{fgr:P_c_all}(b)), \fs{the} simulation results \fs{for the percolation transition} are also higher than the theoretical prediction ($m=\fs{1}$ \fs{corresponding to} $p_Ap_B=1/18$). This discrepancy is consistent with the fact that real associative polymer solutions exhibit significant cycle formation as discussed below in Section \ref{subsec:loop}, which reduces the number of effective connections and thereby shifts the percolation threshold to higher values\cite{wang2017kinetic}. In summary, in both homo- and hetero-associative systems, the theoretical threshold systematically underestimates the critical binding probability required for percolation.

To \fs{better understand the origin of the discrepancies}\fsdel{bridge the gap} between the mean-field prediction and the simulation results, we next analyze the percolation transition using the RG and RGG models introduced in Section~\ref{subsec:graph_theory}.
\fsdel{Since the RG model is independent of spatial embedding and shares the same mean-field assumptions as the \fsdel{Flory--Stockmayer}\fs{F--S} theory, its}\fs{Like the F--S model, the RG model does not account for spatial structure. Cycle formation in finite clusters is not explicitly forbidden, but occurs with vanishing probability if the number of vertices approaches infinity. Therefore, finite clusters effectively have a tree-like structure and the} predicted \fs{percolation} thresholds \fs{in the RG model} coincide\fsdel{exactly} with Eqs.~\eqref{homo_pc} and~\eqref{hetero_pc}. \fsdel{For brevity, these}\fs{To keep the plots simple, the RG} results are presented together with the theoretical predictions. 

In real polymer solutions, bond formation requires two polymers to approach within a finite distance, leading to additional geometric constraints. 
\fsdel{To incorporate such effects, the}\fs{The} RGG model \fs{incorporates such effects by embedding the vertices in space and only allowing bonds between vertices within a cutoff distance, the detection radius $R_\textrm{cut}$} (as shown in Figure~\ref{fgr:flowchart}(b)). \fs{This model} is\fsdel{then} utilized to generate the numerical predictions for the homo- and heteroassociative systems shown in Figure~\ref{fgr:P_c_all} (orange point lines). \fs{Here we have adjusted $R_\textrm{cut}$ manually to obtain optimal agreement between simulations and RGG data.}

\begin{table}
\centering
\caption{
Radius of gyration $R_g$ measured from MD simulations and the corresponding cutoff distance $R_\mr{cut}$ used in the RGG model.
}
\label{tab:rcut_Rg}
\begin{tabular}{lccc}
\hline\hline
 & \multicolumn{1}{c}{Single component} & \multicolumn{2}{c}{Two components} \\
\cline{3-4}
 &  & A & B \\
\hline
$N$        & 65+2$\times$10     & 65     & 23     \\
$R_g/\sigma$      & $4.87\pm0.32$ & $6.48\pm0.09$ & $3.31\pm0.06$ \\
$R_\mr{cut}/\sigma$ 
           & $8\ (\approx 2R_g)$ 
           & \multicolumn{2}{c}{$10\ (\approx R_{g,A}+R_{g,B})$} \\
\hline\hline
\end{tabular}
\end{table}

As shown in Figure~\ref{fgr:P_c_all}, the RGG model provides an accurate description of the percolation threshold in the simulation once an appropriate cutoff distance $R_\mr{cut}$ is chosen. As $R_\mr{cut}$ increases, the gel region expands, reflecting the enhanced connectivity allowed by a larger geometric interaction range. A natural physical hypothesis is \fs{to assume} that effective interchain binding becomes possible when two polymer coils begin to overlap. Motivated by this picture, we compute the average radius of gyration $R_g$ for each polymer species from the simulation trajectories, and summarize the results in Table~\ref{tab:rcut_Rg}. For the homoassociative system, the \fs{radius of gyration for} chain\fs{s} with length $N=65(+2\times10)$ is $R_g\approx4.87\sigma$. The optimized \fs{RGG} cutoff \fs{radius} for the single-component system is $R_{\mathrm{cut}} = 8\sigma$, which is of the order of the typical center-of-mass separation between two coils (approximately $2R_g$). For \fs{the} heteroassociative system, $R_{g,A}+R_{g,B}\simeq 9.79\sigma$, in excellent agreement with the \fs{optimized RGG} cutoff \fs{radius} $R_\mr{cut}=10\sigma$. This supports the physical picture that interchain bonding requires direct coil overlap, so that the sum of the gyration radii naturally defines the relevant contact distance. \fsdel{Overall, the} \fs{The} RGG detection radius can therefore be interpreted as a physically motivated conformational length scale, rather than an arbitrary fitting parameter. 

The thresholds obtained from simulation and from the RGG model nearly coincide over a broad parameter range, particularly in the semidilute regime. This agreement indicates that the geometric constraint encoded in the RGG model captures the dominant spatial effect governing percolation. \xc{However, a small deviation remains in the dilute regime, ($c<c^*\approx0.18\sigma^{-3}$) \fs{in the}\fsdel{for} homoassociative case, as shown in Figure~\ref{fgr:P_c_all}(a). \fs{One possible source of disagreement is that spatial correlations between polymers may become important in this regime, which are neglected in the RGG model. Much more importantly, however, the binding energies \xc{for} percolation at low densities are very high ($\epsilon_{\mr{sp}}\xc{>}13k_BT$), such that the lifetime of bonds become very large ($\tau_b \sim 4 \times 10^5 t_0$ for $\epsilon_{\mr{sp}}=13k_BT$) and the system could not properly equilibrate. \xcc{Figure S2 in the Supporting Information shows the binding lifetimes in the two models.}
In Figure~\ref{fgr:P_c_all}, }closed symbols denote equilibrated configurations, whereas the open symbols represent \fsdel{the} threshold points for which the bond lifetime is comparable to or longer than the entire simulation time and equilibration \fsdel{cannot} \fs{could not} be established   ($\epsilon_{\mr{sp}}\xc{>}13k_BT$). 
We \fsdel{retain} \fs{include} these points \fs{in the figure} to illustrate the crossover toward an effectively irreversibly crosslinked network. \fs{The initial configurations of the simulations were set up as uncrosslinked polymer solutions, and bonds between binding sites progressively formed. At high binding energies, such}\fsdel{. In this regime,} formed bonds rarely reopen, while the polymer chains remain mobile and can encounter additional binding partners. The resulting persistent connections accumulate over time, potentially allowing percolation at lower concentrations than predicted by the static, fixed-configuration RGG \fs{with reversible binding}. This is why the simulation threshold will be lower than the RGG prediction.}


\subsection{Cycle analysis}
\label{subsec:loop}

It is well established that deviations from the \fsdel{Flory--Stockmayer}\fs{F--S} prediction in real polymer systems are due to the formation of cycles within the network\cite{wang2017kinetic}. In particular, cycles do not contribute to cluster expansion, thereby suppressing the growth of the giant component and shifting the percolation boundary. Motivated by these studies, we next quantify cycle formation in both associative systems and identify the dominant microscopic mechanisms.

\begin{figure}[ht]
\includegraphics[width=8cm]{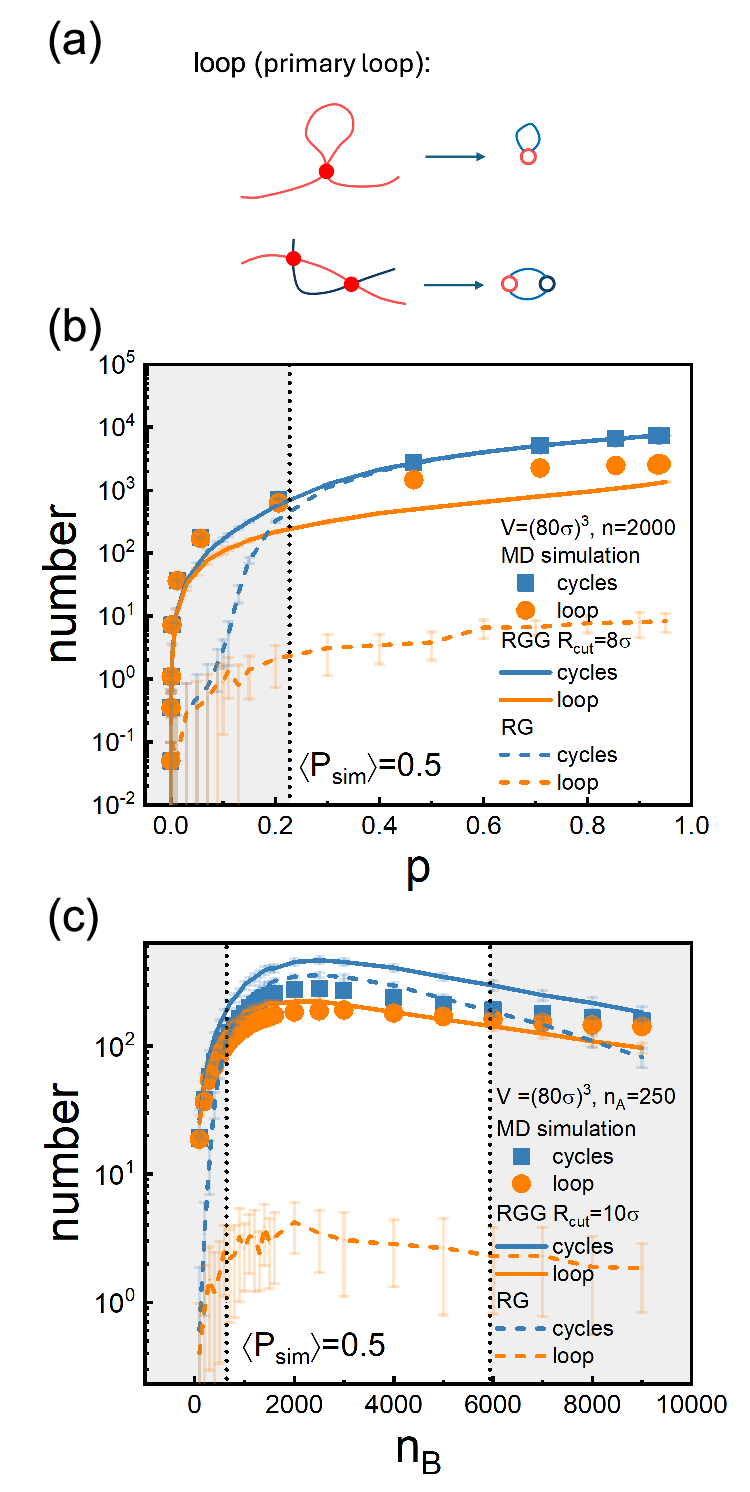}
\caption{(a) Schematic illustration of \fs{a} primary loop in each system: intra-chain binding in \fs{the} homoassociative system, and a minimal cycle in \fs{the} heteroassociative system. (b,c) Comparison of cycle statistics \fsdel{among}\fs{in} MD simulations, RGG, and RG models for homo- and hetero-associative polymer systems, respectively. Symbols denote MD simulation results, solid lines the corresponding RGG prediction with an optimized cutoff $R_{\mr{cut}}$, and dashed lines the RG results. The vertical black dotted lines indicate the percolation threshold defined by $\langle P\rangle=0.5$ in simulation, while the shaded regions highlight the gel-free regime.}
\label{fgr:loop_all}
\end{figure}

As discussed in Section~\ref{subsec:theory}, we distinguish between total cycles and a special class of minimal cycles, which we refer to as primary loops. Specifically, in a homo-system, \fsdel{the}\fs{a} primary loop \fsdel{corresponds to an intra-chain binding where} \fs{is created by an intra-chain bond between} two binding sites on the same polymer chain.\fsdel{ form a bond.} In a hetero-system, \fsdel{the}\fs{a primary} loop is defined as a minimal cycle involving \fsdel{exactly} four crosslinks between two distinct polymer chains. Figure~\ref{fgr:loop_all}(a) illustrates the structure of these two primary loops. The loop statistics are then measured and compared across different models in Figures~\ref{fgr:loop_all}(b,c).

For the homoassociative system(Figure~\ref{fgr:loop_all}(b)), the RGG model yields a markedly more accurate prediction of both cycle and loop numbers than the RG model, reproducing the simulation data across the full range of $p$. \fsdel{Compared with the RG predictions (dashed lines), both}\fs{Both} the MD simulation data and the RGG results exhibit a pronounced excess of primary loops in the pre-gel regime (shaded region), indicating that intrachain bonding constitutes the dominant source of cycle formation prior to percolation. These intrachain loops reduce the effective chain functionality\cite{zhong2016quantifying,zhang2025topology}, providing a natural explanation for the upward shift of the percolation threshold observed in Figure~\ref{fgr:P_c_all}. 

An analogous trend is observed in the heteroassociative system(Figure~\ref{fgr:loop_all}(c)), where the RGG framework again yields improved agreement with MD simulation. Notably, fixing one component while increasing the concentration of the other drives the system through a reentrant percolation transition\cite{danielsen2023phase,chen2025sol}, giving rise to two distinct gel-free regions, as highlighted by the shaded areas in Figure~\ref{fgr:loop_all}(c). In both associative systems, primary loops remain rare in the RG model, consistent with its loop-free mean-field character \fs{in the thermodynamic limit} and hence with its close correspondence to the classical F--S theory. 

Taken together, these results identify primary-loop formation as the \fsdel{key}\fs{dominant} mechanism responsible for the deviation of the percolation threshold from F--S theory: intrachain bonding or minimal cycle \fs{formation} reduces the effective functionality \fs{and the number of bonds contributing to cluster expansion}, systematically shifting the gel point \fsdel{upward}\fs{towards larger binding probabilities}. 
\fsdel{The loop analysis above reveals that a large number of primary loops form already in the pre-gel regime(see Figure~\ref{fgr:loop_all}(b,c)). For the homoassociative or heteroassociative systems, primary loop dominates cycle formation at low concentrations, consuming reactive sites within the same chain without contributing to cluster expansion. This effectively reduces the functionality available for network growth below the intrinsic valence $f$ and effective binding probability.}

\begin{figure}[ht]
\includegraphics[width=8cm]{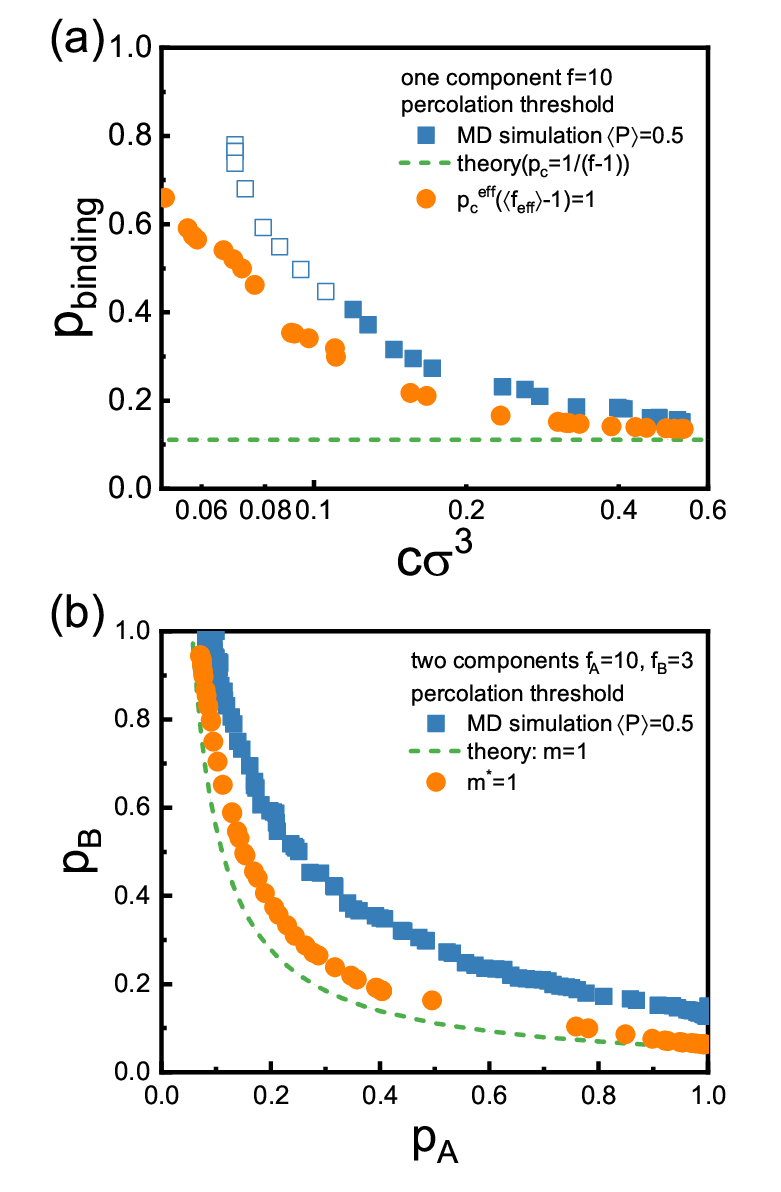}
\caption{Comparison between \fs{percolation thresholds predicted by} the original   and loop-corrected F--S percolation criteria \fs{and MD simulation results (blue squares, same data as Figure~\protect\ref{fgr:P_c_all})}. (a) Homoassociative system: the dashed line denotes the original F--S prediction, ($p_c=1/(f-1)$), and the orange circles the modified \fs{primary loop-corrected} prediction \fsdel{including primary-loop corrections} (Eq.~\eqref{eq:homo_loop_corrected_threshold}). (b) Heteroassociative system: the dashed line denotes the original F--S criterion, ($m=1$) (Eq.~\eqref{hetero_pc}), and the orange circles the corrected boundary from the modified criterion, ($m^\ast=1$) (Eq.~\eqref{eq:hetero_pc_modify}). \fsdel{MD results (blue squares) are shown for reference}.}
\label{fgr:P_c_modify}
\end{figure}

To account for this, we renormalize the F--S percolation criterion by replacing $f$ with an averaged effective functionality $\langle f_{\mr{eff}}\rangle$ \xc{and \fs{$p$ by a} loop-free binding probability $p_\mr{eff}$} (see Eqs.~\eqref{eq:homo_loop_corrected_threshold}\cite{dobrynin2004phase,santra2021universal} and \eqref{eq:hetero_pc_modify}), which quantifies the loss of reactive sites \fs{and the presence of redundant bonds} due to primary loop\fs{s} in the pre-gel regime. For homoassociative system\fs{s}, as shown in Figure~\ref{fgr:P_c_modify}(a), this correction (Eq.~\eqref{eq:homo_loop_corrected_threshold} shifts the predicted threshold upward and significantly improves agreement with simulation, particularly at \fsdel{semi-dilute}\fs{low} concentrations where primary loops dominate the binding structure.

For heteroassociative system\fs{s}, intrachain bonding is forbidden by construction (A chains bind exclusively to B chains). Nevertheless, the loop statistics in Figure~\ref{fgr:loop_all}(c) show that a significant number of primary loops, i.e., minimal cycles involving both species, are present in the pre-gel regime. Applying the analogous renormalization (Eq.~\eqref{eq:hetero_pc_modify}) to the heteroassociative F--S criterion,  the correction is less successful, as shown in Figure~\ref{fgr:P_c_modify}(b). \fsdel{The remaining discrepancy likely arises because a heteroassociative cycle is not a primary loop in the same physical sense as an intrachain loop in homo-case.} \fs{We attribute this difference to the distinct role of primary loops in the two systems. In the homoassociative case, primary loops correspond to intrachain cycles that do not create links between different chains and therefore do not contribute to network connectivity. By contrast, primary loops in the heteroassociative system are second-order cycles that connect different chains, allowing them to increase cluster size in much the same way as higher-order cycles.} Related limitations of \fsdel{some} corrections for nonlocal cycles have\fsdel{also} been discussed in our previous work~\cite{chen2025sol}.


\begin{figure*}[t]
\includegraphics[width=17cm]{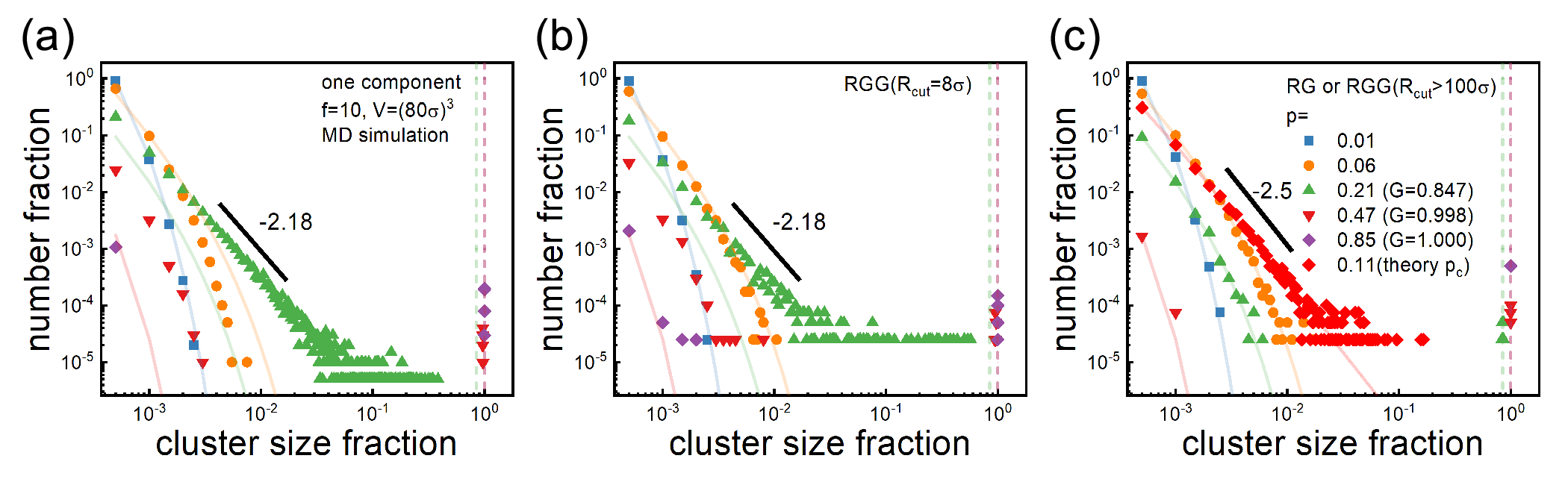}
\caption{Cluster-size distribution as a function of the cluster size fraction for different binding probabilities $p$, obtained from (a) MD simulations, (b) RGG with \fsdel{proper} detection radius $R_\mr{cut}=8\sigma$, and (c) RG or RGG with \fsdel{a large enough}\fs{an extremely large} detection radius($R_\mr{cut}>100\sigma$). Light solid lines indicate the theoretical prediction \xc{(calculated by Eq.~\eqref{eq:ns_single_final})}. Light dashed vertical lines mark the giant-cluster fraction $G=n^\mr{giant\ cluster}_\mr{chain}/n_\mr{chain}$ at the corresponding $p$.}
\label{fgr:homo_size}
\end{figure*}

\subsection{Cluster size distribution and giant cluster}
\label{subsec:cluster_size}

The F--S theory not only predicts the percolation threshold, but also provides analytical predictions for the cluster-size distribution and the giant-cluster fraction under the tree-like assumption\cite{stockmayer1943theory,stockmayer1944theory,stockmayer1952molecular}. Therefore, we further compare the cluster-size distribution obtained from MD simulations with those from RG and RGG models. \fs{In Appendix~\ref{derive_GF}}\fsdel{Here}, we derive explicit expressions \fsdel{of}\fs{for} both the cluster-size distribution and the giant-cluster fraction\cite{tavares2010equilibrium,corezzi2008molecular}, \fsdel{through }\fs{using a} generating-function formalism \fsdel{with}\fs{and} the Lagrange inversion theorem.\fsdel{(see Appendix~\ref{derive_GF}).}

Figure~\ref{fgr:homo_size}(a--c) compares the cluster-size distributions of the homoassociative system obtained from (a) MD simulations, (b) \fs{the} RGG model, and (c) \fs{the} RG model, for several binding probabilities $p$. The MD simulation results (Figure~\ref{fgr:homo_size}(a)) agree with the F--S prediction only in the weak-binding regime ($p < 0.06$). As $p$ increases, the percolation transition in the simulation is \fsdel{delayed}\fs{shifted} relative to the F--S prediction, with the simulated threshold point $p_c \approx 0.21$ higher than the mean-field value $p_c \approx 0.11$. By contrast, \fs{the} RG model (or the RGG model with a sufficiently large detection radius)\fsdel{recovers} \fs{reproduces} the theoretical cluster-size distribution \xc{(see Eq.~\eqref{eq:ns_single_final})} exactly (Figure~\ref{fgr:homo_size}(c)). The \fs{results from the} RGG model with an appropriate detection radius $R_{\mathrm{cut}} \approx 2R_g$ (Figure~\ref{fgr:homo_size}(b)) \fs{are again in good agreement with the MD simulation results.}\fsdel{again reproduces the MD simulation results quantitatively.}

\fsdel{Moreover, beyond}\fs{Beyond} the percolation threshold, the cluster-size distribution in \fs{the} simulation\fs{s} becomes strongly bimodal: \fs{the system is dominated} \fs{it is characterized} by the coexistence of a macroscopic gel cluster (with size fraction approaching $1$) and a population of small clusters. The intermediate-size clusters are strongly depleted\cite{chen2025sol,jedlinska2024effects,ranganathan2020dynamic}. At the percolation threshold ($p = 0.21$ for $\langle P \rangle \approx 0.5$), the distribution \fs{of finite clusters with size $s$ follows a power law}\fsdel{exhibits a well-defined power-law scaling regime}, $n_s \sim s^{-\tau}$, \fs{in all three models. In the MD simulations and in the RGG model, the scaling exponent is}\fsdel{with exponent} $\tau \approx 2.18$, in quantitative agreement with the universal critical exponent of three-dimensional percolation\cite{li2022distribution,corezzi2008molecular,de1977critical,RubinsteinColby}. This is in contrast to the RG model, \fsdel{which follows}\fs{which features} the mean-field Fisher exponent $\tau = 5/2$. \fsdel{characteristic of the universality class. But the RGG model reproduces the three-dimensional percolation exponent $\tau \approx 2.18$. This demonstrates} \fs{The results demonstrate} that the critical cluster statistics near the gel point are governed by the universality class of three-dimensional percolation, and that finite-range spatial constraints \fs{as} encoded in the RGG framework are essential to recover the correct critical behavior\cite{li2022distribution}.

\begin{figure*}[ht]
\includegraphics[width=17cm]{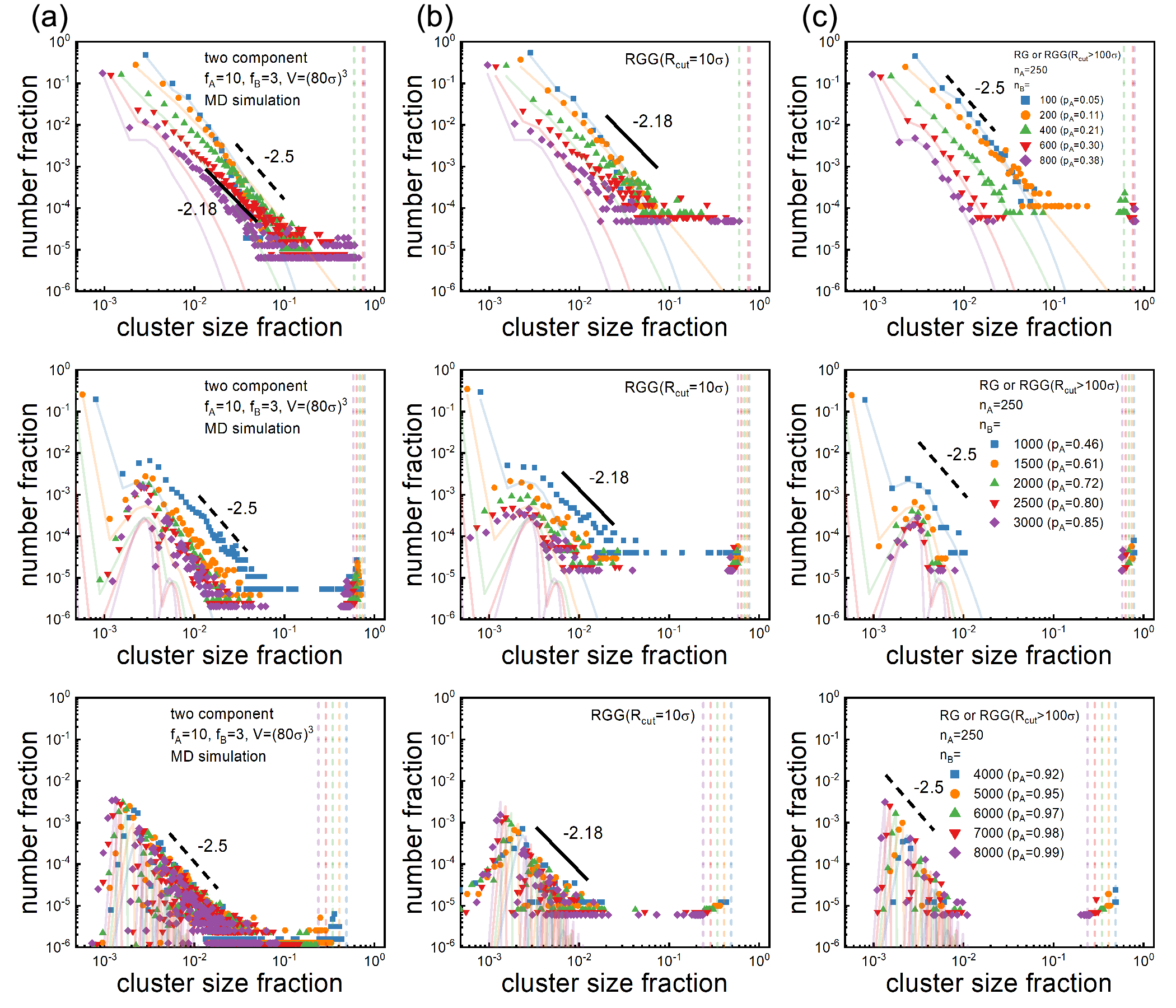}
\caption{Cluster-size distributions of the heteroassociative system as a function of the cluster-size fraction for fixed A-component number ($n_A=250$) and increasing B-component number($n_B$), obtained from (a) MD simulations, (b) RGG with \fsdel{the appropriate} detection radius $R_\mr{cut}=10\sigma$, and (c) RG or RGG with \fsdel{a large enough}\fs{extremely large} detection radius($R_\mr{cut}>100\sigma$). Light solid lines denote the corresponding theoretical predictions \xc{(calculated by Eq.~\eqref{eq:A16})}, and light dashed vertical lines mark the giant-cluster fraction($G=n_\mr{chain}^\mr{giant}/n_\mr{chain}$) at the corresponding $p$, which are listed in Table~\ref{tab:giant_fraction}. From top to bottom, the rows correspond to the pre-gel regime (no percolation), \fsdel{post-}\fs{the }gel regime (percolation present), and \fs{the} re-entrant regime (return to the sol state), respectively.}
\label{fgr:hetero_size}
\end{figure*}

The same picture holds for the heteroassociative system (Figure~\ref{fgr:hetero_size}(a--c), fixed $n_A = 250$, increasing $n_B$): the RG model recovers the F--S prediction, while the RGG model quantitatively reproduces the MD data. Focusing on the MD results for the heteroassociative system(Figure~\ref{fgr:hetero_size}(a)), in the pre-gel regime(top row), the cluster-size distribution decays monotonically with increasing cluster size. As $n_B$ increases, \fs{or} equivalently as the effective binding probability $p_A$ increases at fixed $n_A$, the distribution progressively broadens, signaling the growth of large finite clusters and the approach to percolation. 

For $n_B<650$ (corresponding to the point for $\langle P\rangle=0.5$), the system remains in the sol regime. Near the threshold ($n_B=600$, red points in first row), the distribution exhibits power-law scaling with exponent $\tau \approx 2.18$, again consistent with the universality class of three-dimensional percolation. By contrast, away from the critical regime, the cluster statistics are much closer to the mean-field prediction. Taken together, both the homoassociative and heteroassociative systems exhibit the same crossover: mean-field-like behavior far from the threshold, but three-dimensional universal scaling around the percolation threshold.

In the \fsdel{post-}gel regime (middle row of Figure~\ref{fgr:hetero_size}(a)), a giant cluster coexists with a population of small clusters. As the giant cluster grows with increasing $n_B$, the small-cluster distribution progressively narrows, reflecting the incorporation of isolated components into the giant cluster. Upon further increasing $n_B$, the system enters a reentrant sol regime (bottom row): excessive saturation of hetero-binding sites suppresses long-range connectivity, and the giant cluster dissolves due to stoichiometric over-saturation, consistent with our previous study\cite{chen2025sol}.

A further feature specific to asymmetric heteroassociative systems is revealed by the theoretical predictions in Figure~\ref{fgr:hetero_size}(c). In the \fsdel{post-}gel regime, the cluster-size distribution develops \fsdel{some} local minima at discrete cluster sizes (e.g., $s = 2, 12, 22, \ldots$), a feature also visible in both the MD simulations and the RGG model. These minima have a combinatorial origin: within the F--S framework, every bond consumes one reactive site on an A chain and one on a B chain, \fsdel{so}\fs{therefore} the number of reacted sites must satisfy $p_A n_A f_A = p_B n_B f_B$. When $f_A \neq f_B$, only specific integer combinations of A and B degrees are compatible with this constraint, leading to a combinatorial suppression of certain cluster sizes. This discrete valence-matching effect vanishes when the stoichiometric asymmetry is removed, i.e., when $f_A = f_B$ or $n_A f_A \approx n_B f_B$.

\begin{table}[ht]
\centering
\caption{\xc{Theoretical} giant-cluster fraction for different numbers of B chains at fixed A-chain number ($n_A=250$), together with the corresponding binding probability ($p_A$).}
\label{tab:giant_fraction}
\begin{tabular}{ccc}
\hline\hline
$n_B$ & $p_A$ & Giant-cluster fraction $G$ \\
\hline
100   & 0.05 & $-$ \\
200   & 0.11 & $-$ \\
400   & 0.21 & 0.59 \\
600   & 0.30 & 0.75 \\
800   & 0.38 & 0.78 \\
1000  & 0.46 & 0.78 \\
1500  & 0.61 & 0.74 \\
2000  & 0.72 & 0.69 \\
2500  & 0.80 & 0.63 \\
3000  & 0.85 & 0.58 \\
4000  & 0.92 & 0.49 \\
5000  & 0.95 & 0.41 \\
6000  & 0.97 & 0.34 \\
7000  & 0.98 & 0.29 \\
8000  & 0.99 & 0.24 \\
\hline\hline
\end{tabular}
\end{table}

Regarding the giant-cluster fraction $G=n^\mr{giant}_\mr{chain}/n_\mr{chain}$ (dashed vertical lines in Figures~\ref{fgr:homo_size}  \fs{and \ref{fgr:hetero_size}}), the F--S prediction overestimates $G$ above the gel point $p_c$. This discrepancy is \fsdel{even more}\fs{particularly} pronounced in the heteroassociative system(Figure~\ref{fgr:hetero_size}). Because bond formation is limited by spatial proximity, the growth of the giant cluster is not controlled solely by functionality and binding probability. Even when reactive groups are available, small clusters may remain spatially isolated from larger aggregates and therefore cannot merge efficiently into the giant component. Consequently, the giant-cluster fraction grows more slowly and remains smaller than the mean-field prediction.

\begin{figure*}[ht]
\includegraphics[width=13cm]{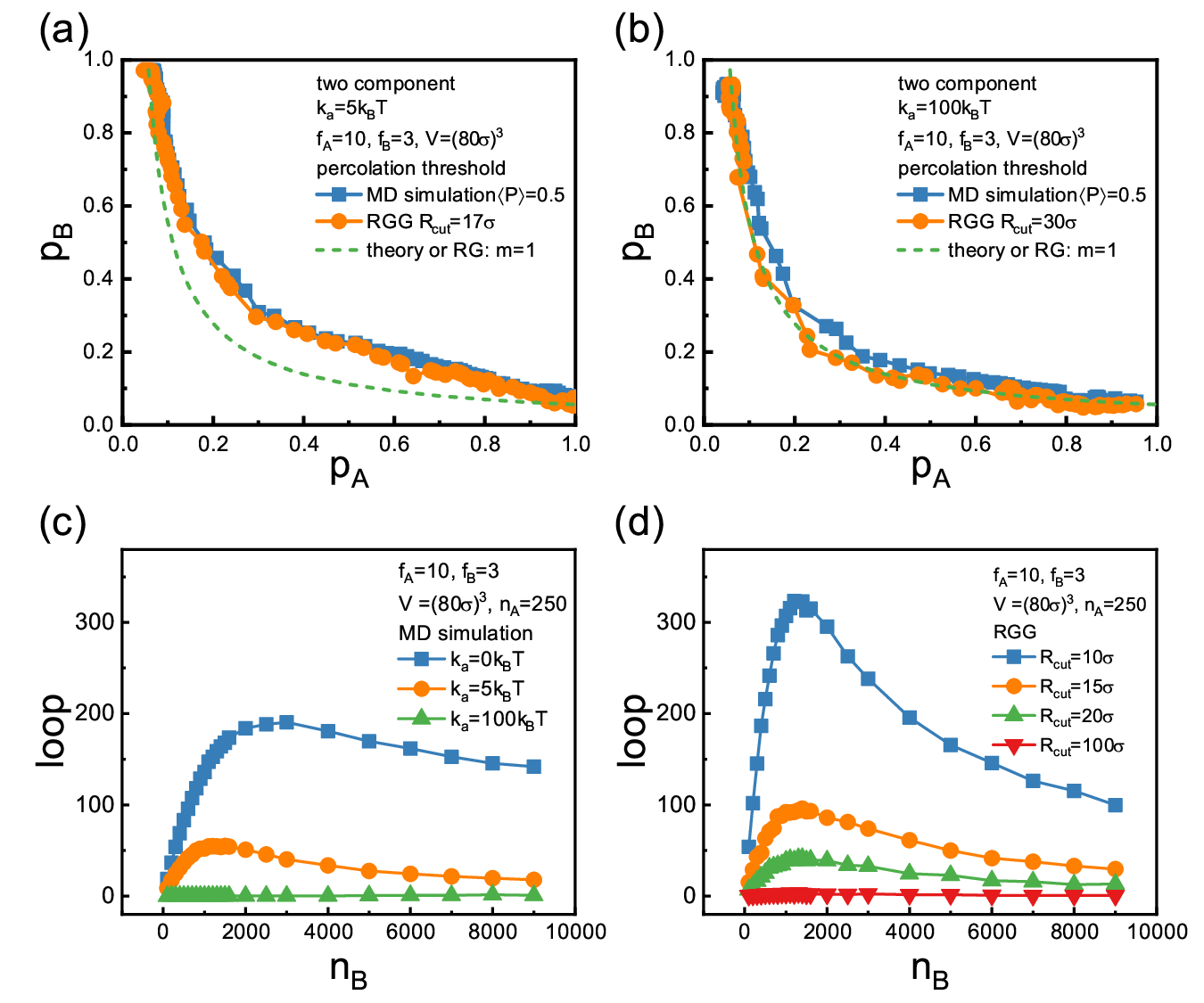}
\caption{Percolation thresholds of the heteroassociative system at different chain stiffnesses, comparing MD simulation results with predictions from the RGG model and the mean-field theory. (a) and (b) correspond to angular potential constants ($k_a=5k_B T$) and ($k_a=100k_B T$), respectively. (c) shows the loop number as a function of stiffness from MD simulations, and (d) the corresponding loop number predicted by the RGG model for different detection radii ($R_\mr{cut}$).}
\label{fgr:stiffenss}
\end{figure*}

\subsection{Stiffness effect on the percolation transition}
\label{subsec:stiffness}

The analyses above establish that the RGG model captures all relevant features \fs{of the homo- and heteroassociative polymer systems} quantitatively, reproducing the simulation results for the cluster-size distribution, the giant-cluster fraction, and the percolation threshold. To further test the robustness of this framework, we examine how chain conformation\fs{s} affect\fsdel{s} \fs{the} network topology and the accuracy of the RGG model. Chain stiffness serves as one of control parameters: stiffer chains suppress intrachain contacts and thereby reduce primary-loop formation (see Figures~\ref{fgr:stiffenss}(c)). A similar suppression has been reported in multivalent patchy particle models, where geometric rigidity effectively eliminates loops and yields excellent agreement with the F--S prediction across a wide range of reaction extents\cite{corezzi2008molecular}. Here, we vary \fs{the} chain stiffness by tuning the angular potential constant $k_a$, and monitor its effect on loop statistics and the percolation threshold. Since the homoassociative and heteroassociative systems display similar behavior, we focus on the heteroassociative case in the following.

\begin{table}
\centering
\caption{
Radius of gyration $R_g$ \fs{of A- and B polymer chains} at different angular potential constants $k_a$ measured from MD simulations for the heteroassociative system and the corresponding cutoff distance $R_\mr{cut}$ used in \fs{the} RGG model.
}
\label{tab:stiff_rg}
\begin{tabular}{c cc c}
\hline\hline
$k_a/k_BT$ 
& \multicolumn{2}{c}{$R_g/\sigma$} 
& $R_\mr{cut}/\sigma$ \\
\cline{2-3}
& $N_A=65$ & $N_B=23$ & $\approx R_{g,A}+R_{g,B}$\\
\hline
$0$ & $6.48\pm0.09$ & $3.31\pm0.06$ & $10$ \\
$5$ & $11.55\pm0.14$ & $5.78\pm0.04$ & $17$ \\
$100$ & $21.81\pm0.02$ & $8.01\pm0.01$ & $30$ \\
\hline\hline
\end{tabular}
\end{table}

As summarized in Table~\ref{tab:stiff_rg}, increasing stiffness leads to an increase of the \fsdel{radius}\fs{radii} of gyration $R_g$, indicating that polymer chains adopt more extended configurations. This conformational expansion reduces the probability of intrachain contacts, thereby suppressing primary-loop formation, and \fsdel{providing}\fs{provides} a direct structural basis for the stiffness-dependent impact of the percolation threshold.

Figures~\ref{fgr:stiffenss}(a,b) compare the percolation thresholds obtained from MD simulations with theoretical predictions for semi-flexible($k_a=5k_BT$) and fully rigid chains($k_a=100k_BT$), respectively. With increasing angular potential constant $k_a$, the simulation results approach the theoretical percolation boundary($m=1$). Notably, the RGG model maintains quantitative agreement with the MD results across all stiffness regimes, without any \fsdel{other}\fs{further} parameterization. This demonstrates that the RGG framework is robust with respect to chain conformation, and that the detection radius $R_{\mathrm{cut}}$, set by the gyration radius $R_g$, remains the only physically motivated length scale needed to accurately locate the percolation threshold.

To elucidate the effect of stiffness, we analyze the loop statistics as a function of $k_a$ in Figure~\ref{fgr:stiffenss}(c). Primary-loop formation is suppressed with increasing stiffness, and is absent for fully rigid chains ($k_a = 100,k_BT$). This trend is consistently captured by the RGG model: as shown in Figure~\ref{fgr:stiffenss}(d), increasing the detection radius $R_{\mathrm{cut}}$ suppresses loop formation in the RGG framework, mirroring the effect of chain stiffness in simulation. In the rigid limit, chains are fully extended and the effective detection radius becomes comparable to the system size, so that each binding site can interact with nearly all other sites and spatial constraint becomes negligible. The network topology therefore converges to the tree-like limit assumed in classical mean-field theory, recovering near-perfect agreement between simulation and the F--S prediction, as shown in Figure~\ref{fgr:stiffenss}(b).

\section{Summary}

In this work, we investigated the percolation transition in homoassociative and heteroassociative polymer solutions using MD simulations together with random graph and random geometric graph models. Our main results can be summarized as follows:

\begin{itemize}
  \item We constructed percolation phase diagrams based on the equilibrium binding probability, i.e., ($c$, $p$) for \fsdel{the} homoassociative polymer system\fs{s} and ($p_A$, $p_B$) for \fsdel{the} heteroassociative two-component system\fs{s}. \fsdel{Two}\fs{Different} simulation models \fsdel{are}\fs{were} \fsdel{employed}\fs{developed} to achieve one-to-one binding \fs{in the two types of systems}. In both cases, the classical Flory--Stockmayer \fs{(F--S)} theory \fs{was found to} systematically underestimate\fsdel{s} the percolation threshold \fs{compared to simulations}, particularly in dilute regimes. This deviation originates from the abundance of primary loops in the pre-gel regime. Renormalizing the F--S criterion with an effective functionality $f_{\mr{eff}} = (1 - p_{\mr{loop}})f$ accounts for this reduction and improves \fs{the} agreement with the simulated percolation thresholds.\fsdel{in both systems.}
  
  \item Graph-based descriptions clarify the role of spatial constraints by separating topology from geometry. The coordinate-free random graph \fs{(RG)} model reproduces the F--S percolation threshold and cluster-size statistics exactly, confirming that F--S theory corresponds to the loop-free branching limit of an unconstrained random process. In contrast, the random geometric graph \fs{(RGG)} model incorporates a finite interaction range and thus captures the dominant spatial constraint governing bond formation. The RGG predictions quantitatively meet the simulation results over a wide parameter range. These findings indicate that realistic associative polymer systems can be understood as stochastic branching processes constrained by finite-range geometry.
  
  \item For the random geometric graph model, the detection radius carries a clear physical interpretation. The optimal cutoff \fsdel{is}\fs{can be} determined directly by polymer conformations measured in simulation, rather than treated as a fitting parameter: $R_{\mr{cut}} \approx 2R_g$ for the homoassociative system and $R_{\mr{cut}} \approx R_{g,A} + R_{g,B}$ for the heteroassociative system, establishing a direct link between the percolation threshold and the microscopic chain geometry. This correspondence remains robust upon varying chain stiffness: updating $R_{\mr{cut}}$ according to the stiffness-dependent $R_g$ is sufficient to maintain the agreement with simulation across all stiffness regimes.
  
  \item \fs{An analysis}\fsdel{Analysis} of the cluster-size distribution and giant-cluster fraction, derived via generating functions and the Lagrange inversion theorem, shows that the RG model recovers the mean-field F--S distributions exactly, as expected from its loop-free construction. By contrast, both MD simulations and the RGG model exhibit broadened cluster-size distributions and a smaller giant-cluster fraction relative to the mean-field prediction. Notably, at the gel point, the cluster-size distribution displays power-law scaling consistent with the universality class of three-dimensional percolation, rather than the mean-field Fisher exponent, confirming that finite-range spatial constraints are responsible for the crossover between universality classes.
\end{itemize}

While the present work establishes a mapping between associative linear polymer systems and random geometric graphs,  several important questions remain open. First, our validation has been carried out exclusively for linear chain architectures with uniformly distributed binding sites along the backbone. In this case, the spatial extent of a chain is well characterized by a single length scale, $R_g$, which directly sets the RGG detection radius. However, when binding sites are distributed non-uniformly, for instance, when stickers are clustered into large blockness\cite{rasid2021effect,qin2025binding,chen2026dilute}, the local binding-site accessibility and intrachain correlations are fundamentally altered. Such block-like architectures may introduce additional length scales and stronger spatial correlations that are not captured by a single $R_{\mr{cut}}$, potentially requiring an extended RGG description. Second, it is not yet clear whether the same correspondence holds for branched or star polymer architectures, where the local topology and binding-site geometry differ fundamentally from the linear case. Extending \fsdel{this} \fs{the present} framework to both non-uniform sticker distributions and branched polymer systems will be \fsdel{addressed in}\fs{an interesting project for} future work.

Furthermore, the mapping developed in this work is \fsdel{most directly applicable} \fs{restricted} to the reversible crosslinking regime, where the binding strength is sufficiently weak that bond formation and breaking occur frequently on \fsdel{the time scale of structural relaxation}\fs{the accessible time scales}. \fs{In such cases, the network topology is annealed and solely depends on chain architectures and binding energies. In practice, it will also depend on the history of the material and adapt over time. It will be interesting to test whether the RGG framework (possibly with dynamically moving vertices) can also provide insights into the topology of such nonequilibrium networks. To address such problems, a RGG mapping would also require a careful calibration of vertex dynamics and binding kinetics, possibly including memory effects arising from cooperative binding or from topological constraints that accumulate during the network formation\cite{qin2025binding}.}
\fsdel{In this limit, the binding events are essentially independent and carry no significant memory of prior network configurations. This memoryless character is the condition in which the reversible binding process can be mapped to an RGG-type stochastic geometric process, and it is well satisfied in the systems considered here. It should be emphasized that, although binding kinetics do not alter the true equilibrium state, sufficiently rapid bond exchange and conformational relaxation are required for the equilibrium ensemble to be sampled within the accessible timescale. 
However, the memory effects may emerge when binding becomes stronger or more complex due to sequence-specific interactions, cooperative binding or topological constraints that accumulate over network formation\cite{qin2025binding}. In such cases, the validity of the RGG mapping may break down, and a more refined model accounting for correlated bond formation would be required. }

Nevertheless, the quantitative agreement between RGG predictions and molecular dynamics simulations considered here demonstrates that \fsdel{this}\fs{RGG} mapping provides a direct and efficient framework for predicting gelation behavior from microscopic chain properties. We therefore expect that the present approach will serve as a useful foundation for understanding network formation in a broad class of associative polymer systems.


\begin{acknowledgments}

XC thanks Supriyo Naskar for useful discussions.
This research was supported by the German Science Foundation (DFG) - Project number 464588647 -- via SFB 1551 (project R05, Subroject number 518287983).
The authors gratefully acknowledge the computing time provided to them on the high-performance computer
Mogon2 and Mogon NHR South-West. 

\end{acknowledgments}


\section*{Supporting Information}

\setcounter{section}{0}
\renewcommand{\thesection}{S\arabic{section}}

\setcounter{figure}{0}
\renewcommand{\thefigure}{S\arabic{figure}}

\section{Relation between binding probability and binding strength}

\begin{figure}[h!]
\centering
\includegraphics[width=7cm]{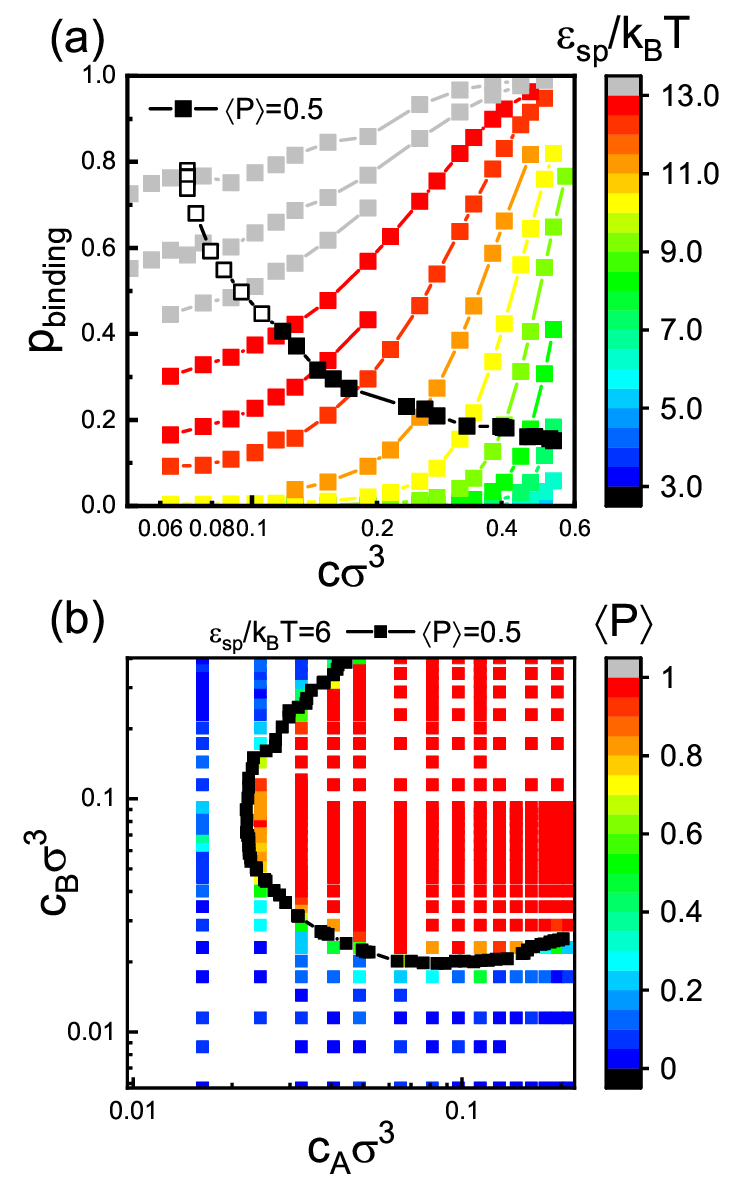}
\caption{(a) 
Binding probability vs. monomer concentration for different binding strengths $\epsilon_\mr{{sp}}/k_B T = 16, 14, 13.5, 13, 12.5, 12, 11, 10, 9, 8, 7, 6, 5$ (from top to bottom, see color coding).
The black squares indicate the percolation threshold defined by $\langle P\rangle=0.5$. Open squares indicate that the binding lifetime at percolation exceeds $0.5 \times10^5 t_0$ (b) Phase diagram for heteroassociative case in the ($c_\mr{A}$,$c_\mr{B}$) plane at $\epsilon_{\mathrm{sp}}/k_{\mathrm{B}}T=6$, colored according to the mean percolation probability $\langle P\rangle$. The black squares mark the percolation threshold, defined by $\langle P\rangle=0.5$. Panel (b) shows the same data as Figure 4 of our previous work \cite{chen2025sol} in different representation. 
}
\label{snapshot}
\end{figure}

\section{Bond lifetimes for the two simulation models}

\begin{figure}[ht]
\centering
\includegraphics[width=8cm]{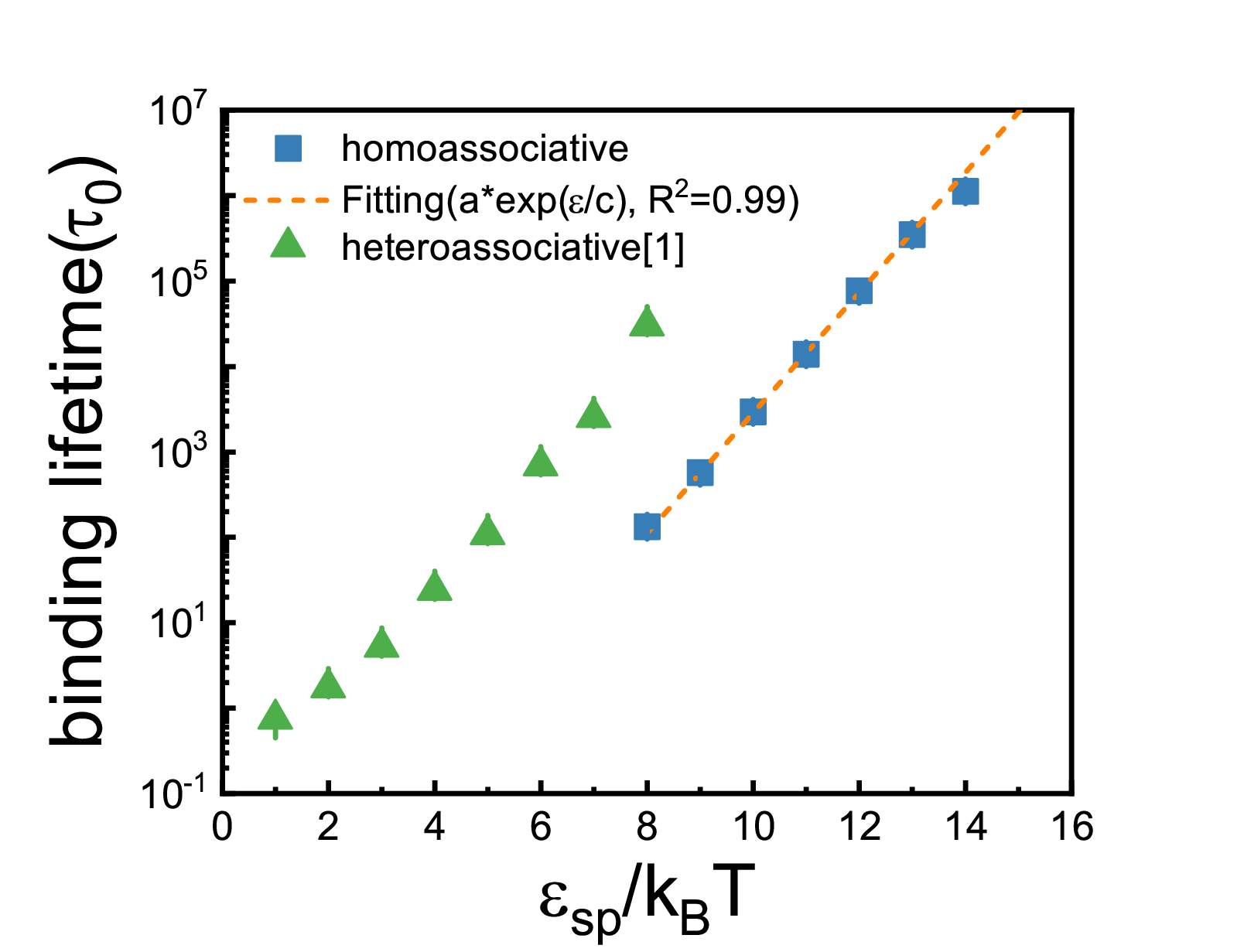}
\caption{Lifetime of reversible binding between two binding domains in the homo- and heteroassociative systems. Blue squares represent the homoassociative system, and the orange dashed line shows a fit to $a e^{\epsilon_\mr{sp}/c}$, which is extrapolated to larger $\epsilon_\mr{sp}$. Green triangles represent the heteroassociative system, which are reproduced from Figure 2 of our previous work \cite{chen2025sol}. The error bars are smaller than the symbols.
}
\label{snapshot}
\end{figure}

\section*{Conflict of Interest}
The authors declare no competing interests.

\section*{Autor Contributions}
{\bf Xinxiang Chen}: Conceptualization (equal); Methodology (equal); Software (lead); Investigation (lead); Formal Analysis (lead); Data Curation (lead); Writing - Original Draft (lead); Writing
- Review and Editing (equal). 
{\bf Lennart Hebestreit}: Software (supporting); Investigation (supporting); Formal Analysis (supporting); Data Curation (supporting) 
{\bf Friederike Schmid}: Conceptualization (equal); Methodology (equal); Resources (lead); Writing - Review and Editing (equal); Supervision (lead); Project Administration (lead); Funding Acquisition (lead).

\section*{Data Availability}
The data supporting the findings of this paper are openly available in Zenodo at \url{https://doi.org/10.5281/zenodo.21609715}, reference number 21609715. The HOOMD scripts used for MD simulations and the source codes for analysis and RG/RGG simulations are available on GitHub: \url{https://github.com/XinxiangChen-git/Hoomd_RG_RGG.git}.

\appendix

\section{Generating-function expressions for cluster statistics}
\label{derive_GF}

\fs{Our goal is to derive analytical expressions for the cluster size distributions of homoassociative and heteroassociative polymer systems in the mean-field case of tree-like graphs. Stockmayer derived such expressions for homoassociative systems in 1943 using combinatorial arguments\cite{stockmayer1943theory} and later supplied, without proof, corresponding expressions for heteroassociative systems\cite{stockmayer1952molecular}. Here, we exploit the generating-function formalism\cite{gordon1962good} to rederive the distributions in a concise manner. }

\subsection{Homoassociative system}

We begin by\fsdel{deriving the cluster size distribution for} \fs{considering} a system of $N$ identical monomers, each containing $f$ functional groups ($f \geq 2$). The extent of reaction is characterized by the conversion $p$ ($0 \leq p < 1$), representing the fraction of functional groups that have reacted. We invoke the standard assumptions of Flory--Stockmayer (F--S) theory: (i) all functional groups of the same type are equally reactive, (ii) reactions occur randomly, and (iii) intramolecular cyclization within finite clusters is neglected. \fsdel{To obtain the cluster size distribution, Stockmayer provided a general expression for this distribution in 1952 \cite{stockmayer1952molecular} without proof. However, since the original derivation was not given explicitly and subsequent works have typically adopted the result without further derivation, it is useful to revisit the formula in more detail. Here, we briefly derive the distribution using the generating-function formulation \cite{gordon1962good}.}

\paragraph{Generating function formulation}

We define the \fsdel{monomer}\fs{molecule}-weighted cluster size distribution $P_s$, normalized as
\begin{equation}
\sum_{s=1}^{\infty} \,P_s \fs{+ G} = 1,
\end{equation}
so that $P_s$ represents the probability that a randomly selected \fsdel{monomer}\fs{molecule} belongs to a \fs{finite} cluster of size $s$.\fs{and $G$ is the fraction of molecules that are part of a giant cluster.}

To characterize the cluster statistics, we introduce two generating functions following the standard percolation and polymer network formalism. The first, $H_0(z)$, generates the probability distribution of cluster sizes observed from a randomly chosen \fsdel{monomer}\fs{molecule}:
\begin{equation}
H_0(z) = \sum_{s=1}^{\infty} P_s z^s.
\label{eq:H0}
\end{equation}
The second, $H_1(z)$, generates the size distribution of finite clusters as viewed from a reacted bond—that is, from a \fsdel{monomer}\fs{molecule} reached by following a randomly chosen bond:
\begin{equation}
H_1(z) = \sum_{s=1}^{\infty} Q_s z^s,
\label{eq:H1}
\end{equation}
where $Q_s$ is the probability that following a bond leads to a finite cluster containing $s$ \fsdel{monomer}\fs{molecule}s.

\paragraph{Recursive relations from bonding statistics}

Consider a \fsdel{monomer}\fs{molecule} reached by following a reacted bond. One of its $f$ functional groups has already been used in the incoming connection, leaving $f-1$ available groups. Each remaining group can be:
\begin{itemize}
\item unreacted, with probability $(1-p)$,
\item reacted, with probability $p$, leading to a subcluster whose size distribution is described by $H_1(z)$. The variables $z$ counts the numbers of \fsdel{monomer}\fs{molecule}s in a cluster.
\end{itemize}
The generating function for one functional group is thus
\begin{equation}
G_{\text{single}}(z) = (1-p)\cdot 1 + p \cdot H_1(z) = 1 - p + p H_1(z).
\end{equation}
Because the $f-1$ branches are statistically independent, the joint generating function for all outgoing branches is
\begin{equation}
G_{\text{total}}(z) = [1 - p + p H_1(z)]^{f-1}.
\end{equation}
Including the central \fsdel{monomer}\fs{molecule} itself (contributing one factor of $z$) gives the fundamental recursion relation:
\begin{equation}
\label{eq:H1_recursion}
H_1(z) = z\,[1 - p + p H_1(z)]^{f-1}.
\end{equation}

For a randomly chosen \fsdel{monomer}\fs{molecule}, all $f$ functional groups are equivalent, leading to
\begin{equation}
\label{eq:H0_recursion}
H_0(z) = z\,[1 - p + p H_1(z)]^{f}.
\end{equation}
Equations~\eqref{eq:H1_recursion} and \eqref{eq:H0_recursion} are the standard recursive relations in F-S theory for finite clusters in the single-component case.

\paragraph{Application of the Lagrange--B\"urmann inversion}

We now extract the coefficients $P_s = [z^s]H_0(z)$ using the Lagrange--B\"urmann formula\cite{stanley2011enumerative,wilf2005generatingfunctionology}.
Equation~\eqref{eq:H1_recursion} can be rewritten as
\begin{equation}
u = z\,\phi(u), \qquad \text{where } \phi(u) = [1 - p + p u]^{f-1},
\end{equation}
and $u \equiv H_1(z)$.  
For any analytic function $F(u)$, the Lagrange--B\"urmann theorem gives
\begin{equation}
[z^n]F(u) = \frac{1}{n}[u^{n-1}]\{F'(u)\,\phi(u)^n\}.
\end{equation}
Substituting $F(u) = [1 - p + p u]^f$, we find
\begin{align}
P_s =& [z^s]H_0(z) = [z^{s-1}] [1 - p + p u]^f \nonumber\\
 =& \fs{\frac{f p}{s-1}[u^{s-2}] [1 - p + p u]^{s(f-1)}}\nonumber\\
 =& \fs{\frac{f}{s-1}\binom{s(f-1)}{s-2}p^{s-1}(1-p)^{sf - 2s + 2},}
\end{align}
\fs{where we have used the binomial theorem in the last step.}
The cluster number distribution (number of clusters of size $s$ per \fsdel{monomer}\fs{molecule}) is given by
\begin{equation}
n_s = \frac{P_s}{s}
= \frac{f}{s(s-1)}\binom{s(f-1)}{s-2}p^{s-1}(1-p)^{sf - 2s + 2}.
\label{eq:ns_single_final}
\end{equation}
Equation~\eqref{eq:ns_single_final} is precisely the F-S cluster size distribution for a single-component system in the mean-field (tree-like) limit.

\paragraph{Giant cluster fraction}

Setting $z=1$ in \fsdel{Eqs.~\eqref{eq:H1_recursion} and \eqref{eq:H0_recursion}}\fs{Eqs.~\eqref{eq:H1} and \eqref{eq:H1_recursion}} yields \fsdel{the self-consistency condition} \fs{a self-consistent equation} for the probability \fsdel{$u$}\fs{$u^*=H_1(1)$} that a randomly chosen bond leads to a finite cluster:
\begin{equation}
u^* = [1 - p + p u^*]^{f-1}.
\label{eq:condition}
\end{equation}
The probability that a randomly selected \fsdel{monomer}\fs{molecule} belongs to a finite cluster (the sol fraction) is
\begin{equation}
S_{\text{sol}} = H_0(1) = [1 - p + p u^*]^f,
\end{equation}
and the gel (giant cluster) fraction is
\begin{equation}
G = 1 - S_{\text{sol}} = 1 - [1 - p + p u^*]^f.
\label{eq:G_gel}
\end{equation}
The gel point corresponds to the onset of a nontrivial solution \fs{of \eqref{eq:condition}} with $u^*<1$, indicating that a reacted bond has a finite probability of connecting to \fsdel{the}\fs{an} infinite cluster. Expanding Eq.~\fsdel{(\ref{eq:G_gel})}\fs{\eqref{eq:condition}} around $u^*=1$ gives the classical \fsdel{Flory--Stockmayer}\fs{F--S} criterion\cite{flory1953principles,stockmayer1943theory}
\begin{equation}
(f-1)p_c=1.
\end{equation}



\subsection{Heteroassociative system}

 We next consider a binary system composed of $N_A$ \fsdel{monomer}\fs{molecule}s of type A with functionality $f_A$ and $N_B$ monomers of type B with functionality $f_B$, where only A--B bonds are allowed. Since every bond consumes one reactive group on each species, the reaction probabilities $p_A$ and $p_B$ are constrained by the stoichiometric relation
 \begin{equation}
 p_A f_A N_A = p_B f_B N_B =: N_p.
 \end{equation}

To describe the finite-cluster statistics, we introduce two generating functions, $H_{1}^{(A)}(z_A,z_B)$ and $H_{1}^{(A)}(z_A,z_B)$
\fs{
\begin{align}
H_{1}^{(A)}(z_A,z_B) &= \sum_{s_A, s_B}^\infty Q_{s_A,s_B}^{(A)} z_A^{s_A} z_B^{s_B}
 \\
H_{1}^{(B)}(z_A,z_B) &= \sum_{s_A, s_B}^\infty Q_{s_A,s_B}^{(B)} z_A^{s_A} z_B^{s_B}
\end{align}
}
for the \fsdel{size distribution of a finite cluster reached by}\fs{probabilities 
$Q_{s_A,s_B}^{(A)}$ and $Q_{s_A,s_B}^{(B)}$ of reaching a finite cluster containing $s_A$ molecules of type $A$ and $s_B$ molecules of type $B$ after} following a randomly chosen reacted bond to an A or B \fsdel{monomer}\fs{molecule}, respectively.  Upon arriving at an A \fsdel{monomer}\fs{molecule} through a reacted bond, one of its $f_A$ reactive groups is already occupied, leaving $f_A-1$ groups available for further branching. Each of these remaining groups is either unreacted with probability $1-p_A$ or connected to a finite branch terminating at a B \fsdel{monomer}\fs{molecule} with probability $p_A H_{1}^{(B)}(z_A,z_B)$. Therefore,
 \begin{equation}
 H_{1}^{(A)}(z_A,z_B)=z_A[1-p_A+p_AH_{1}^{(B)}(z_A,z_B)]^{f_A-1}.
 \label{eq:A9}
 \end{equation}
 By the same argument, for a branch terminating at a B monomer,
 \begin{equation}
 H_{1}^{(B)}(z_A,z_B)=z_B[1-p_B+p_BH_{1}^{(A)}(z_A,z_B)]^{f_B-1}.
 \label{eq:A10}
 \end{equation}
 
For a randomly chosen \fsdel{monomer}\fs{molecule}, all reactive groups contribute independently in the same way. The generating functions for the finite cluster containing a randomly chosen A or B \fsdel{monomer}\fs{molecule} are thus
 \begin{align}
H_{0}^{(A)}(z_A,z_B) =& z_A[1-p_A+p_A H_{1}^{(B)}(z_A,z_B)]^{f_A},
\label{eq:H1A_recursion}\\
H_{0}^{(B)}(z_A,z_B) =& z_B[1-p_B+p_B H_{1}^{(A)}(z_A,z_B)]^{f_B}.
 \label{eq:H1B_recursion}
 \end{align}
Expanding $H_{0}^{(A)}$ and $H_{0}^{(B)}$ in powers of $z_A$ and $z_B$ defines 
the \fsdel{monomer}\fs{molecule}-weighted cluster probabilities $P^{(A)}_{m,n}$ and $P^{(B)}_{m,n}$, i.e., the probabilities that a randomly chosen A or B \fsdel{monomer}\fs{molecule} resides in a finite cluster containing $m$ A \fsdel{monomer}\fs{molecule}s and $n$ B \fsdel{monomer}\fs{molecule}s.

\fs{To obtain explicit expressions for  $P^{(A)}_{m,n}$ and $P^{(B)}_{m,n}$, we apply the multivariate Lagrange-Good inversion formula, which is the higher dimensional extension of the Lagrange-B\"urmann equation\cite{bender1998multivariate}.    }
Let
\begin{equation}
u \equiv H_{1}^{(A)}(z_A,z_B), \qquad v \equiv H_{1}^{(B)}(z_A,z_B),
\end{equation}
which satisfy 
\begin{equation}
u = z_A\,\phi_A(u,v), \qquad v = z_B\,\phi_B(u,v),
\end{equation}
where (see Eqs.~(\ref{eq:H1A_recursion}), (\ref{eq:H1B_recursion})) 
\begin{equation}
\begin{split}
\phi_A(u,v) &\equiv   [1-p_A+p_A v]^{f_A-1} =: G_1^{(A)}(v),\\
\phi_B(u,v) &\equiv   [1-p_B+p_B u]^{f_B-1} =: G_1^{(B)}(u).
\end{split}
\end{equation}

The determinant form of the Lagrange--Good inversion theorem states that for any analytic function $F(u,v)$, the coefficients of a Taylor series satisfy

\begin{eqnarray}
\label{eq:LG_det}
\lefteqn{[z_A^{m} z_B^{n}]\,F(u,v)=} \quad
\\ \nonumber
&&[u^{m} v^{n}]\Big\{
F(u,v)\,
\phi_A(u,v)^{m}\,
\phi_B(u,v)^{n}\,
\Delta(u,v)
\Big\},
\end{eqnarray}
where
\begin{equation}
\Delta(u,v)
=
\det
\begin{pmatrix}
1 - u\,\dfrac{1}{\phi_A}\dfrac{\partial \phi_A}{\partial u} 
&
- u\,\dfrac{1}{\phi_B}\dfrac{\partial \phi_B}{\partial u} \\[2mm]
- v\,\dfrac{1}{\phi_A}\dfrac{\partial \phi_A}{\partial v} 
&
1 - v\,\dfrac{1}{\phi_B}\dfrac{\partial \phi_B}{\partial v}
\end{pmatrix}.
\end{equation}

Applying Eq.~\eqref{eq:LG_det} to $F(u,v)=H_{0}^{(A)}= z_A G_0^{(A)}(v)$ 
with $G_0^{(A)}(v) = (1- p_A + p_A u)^{f_A}$ gives
\begin{equation}
\begin{split}
P_{m,n}^{(A)}=&[z_A^{m-1} z_B^{n}]\,G_{0A}(v)\\
=& [u^{m-1}v^{n}]\,G_{0A}(v)\,G_{1A}(v)^{m-1}G_{1B}(u)^{n}\\
&\times\left[
1 - u v 
\frac{G_{1A}'(v)}{G_{1A}(v)}
\frac{G_{1B}'(u)}{G_{1B}(u)}
\right].
\label{eq:HmnA_coeff}
\end{split}
\end{equation}
An analogous expression holds for $P_{m,n}^{(B)}$ by interchanging $A$ and $B$.

 Carrying out the coefficient extraction via binomial expansion yields the
explicit factorial forms
\fs{
\begin{equation}
P_{m,n}^{(A)} = f_A p_A m \: P_{m,n}^{(0)}, \quad 
P_{m,n}^{(B)} = f_B p_B n \: P_{m,n}^{(0)}
\end{equation}
with
\begin{equation}
 \begin{aligned}
 P^{(0)}_{m,n}
 &=\frac{p_A^{n-1}(1-p_A)^{f_A m -m-n+1} \: (f_A m-m)! }
 {m! \: (f_A m-m-n+1)!}
    \\
 &\times \frac{p_B^{m-1}(1-p_A)^{f_B n -m-n+1}\:  (f_B n-n)! }
 {n! \: (f_B n-m-n+1)!}
 \end{aligned}
 \label{eq:A13}
 \end{equation}
}


The corresponding cluster distribution, $N_{m,n}$, i.e., the number of clusters containing $m$ A \fsdel{monomer}\fs{molecule}s and $n$ B \fsdel{monomer}\fs{molecule}s, is obtained by dividing by the number of \fsdel{monomer}\fs{molecule}s of the selected species in the cluster,
\begin{equation}
N_{m,n}=N_A\frac{P^{(A)}_{m,n}}{m}
       =N_B\frac{P^{(B)}_{m,n}}{n}
       =N_p P_{m,n}^{(0)}.
 \label{eq:A15}
 \end{equation}
 This expression \fsdel{recovers}\fs{reproduces} Stockmayer's molecular cluster distribution for hetero-associative tree-like clusters\cite{stockmayer1952molecular}. Furthermore, the cluster number distribution with respect to the total cluster size $s=m+n$ is \fsdel{then} given by
 \begin{equation}
 n_s=\sum_{m+n=s} N_{m,n}.
 \label{eq:A16}
 \end{equation}

The giant-cluster fraction is determined from the fixed-point equations, ~\eqref{eq:H1A_recursion} and \eqref{eq:H1B_recursion}, by setting $z_A=z_B=1$, in direct analogy with the single-component case. The gel point corresponds to the loss of stability of the trivial finite-cluster solution and is given by the classical hetero-associative Flory--Stockmayer condition\cite{chen2025sol}
 \begin{equation}
 p_Ap_B(f_A-1)(f_B-1)=1.
 \label{eq:A17}
 \end{equation}

\section{Generalised percolation threshold for a polydisperse system.}
\label{derive_pc_generalized}

We derive the generalized threshold conditions using the same recursive framework introduced by Macosko and Miller\cite{macosko1976new}. In their original treatment, the recursion is formulated for the expected branch weight $E(W)$ using the law of total conditional expectation,
\begin{equation}
E(W)=E(W|O)P(O)+E(W|\bar O)P(\bar O),
\end{equation}
where $O$ is an event, $\bar O$ is its complement, and $E(W|A)$ is the conditional expectation of the branch weight $W$ given that A occurs. In the present context, $W$ denotes the total molecular weight contribution associated with a branch reached by following a randomly chosen reactive group. Thus, the expected branch weight can be written recursively by conditioning on whether that group is reacted and, if reacted, on the type of junction molecule reached. We extend this argument to polydisperse associative systems by explicitly accounting for the functionality distribution while assuming identical reaction probabilities for all reactive groups. As in the original Macosko--Miller treatment, the gel point is identified from the divergence of the weight-average molecular weight $M_w$.

\subsection{Homoassociative system}

For a polydisperse homoassociative system containing $n_{f_i}$ molecules \fsdel{of component A }with functionality $f_i$, the probability of reaching a molecule of type $i$ \fsdel{by} following a reacted bond is proportional to the number of reactive groups carried by that species,
\begin{equation}
x_i=\frac{f_i n_{f_i}}{\sum_j f_j n_{f_j}}.
\end{equation}
We denote by $E(W^{\mr{out}})$ the expected branch weight obtained by following a randomly chosen reactive group. If that group is unreacted, the branch stops and contributes zero weight. If it is reacted, it leads to an A molecule of type $i$ with probability $x_i$, and the corresponding expected weight is $E(W_{f_i}^{\mr{in}})$. Thus,
\begin{equation}
E(W^{\mathrm{out}})=0(1-p)+p\sum_i x_i\,E(W_{f_i}^{\mathrm{in}}).
\end{equation}

Once a junction molecule \fsdel{of type A} with functionality $f_i$ is reached, the expected weight looking into that junction is given by the molecular weight of the junction itself plus the expected contributions from its remaining $f_i-1$ branches:
\begin{equation}
E(W_{f_i}^{\mathrm{in}})=M_{f_i}+(f_i-1)E(W^{\mathrm{out}}),
\end{equation}
where $M_{f_i}$ is the molecular weight of \fsdel{component A}\fs{molecules} with functionality $f_i$. Combining these two relations gives
\begin{equation}
E(W^{\mathrm{out}})
=
\frac{p\,m_a}{1-p(\bar f-1)},
\end{equation}
with
\begin{equation}
\bar f=\sum_i x_i f_i,
\quad
m_a=\sum_i x_i M_{f_i}.
\end{equation}
Accordingly,
\begin{equation}
E(W_{f_i}^{\mathrm{in}})
=
M_{f_i}
+
(f_i-1)\frac{p\,m_a}{1-p(\bar f-1)}.
\end{equation}

The expected molecular weight associated with a molecule of type $i$ is then
\begin{equation}
\begin{aligned}
E(W_{f_i})&=E(W^{\mathrm{out}})+E(W_{f_i}^{\mathrm{in}})\\
&=M_{f_i}+f_i E(W^{\mathrm{out}}).
\end{aligned}
\end{equation}

The weight-average molecular weight is obtained by averaging over all molecules with weight fraction
\begin{equation}
w_{f_i}=\frac{M_{f_i}n_{f_i}}{\sum_i M_{f_i}n_{f_i}},
\end{equation}
so that
\begin{equation}
M_w=E(W)=\sum_i w_{f_i}E(W_{f_i}).
\end{equation}
Substituting the above expressions gives
\begin{equation}
M_w
=
\sum_i w_{f_i}M_{f_i}
+
\sum_i w_{f_i}f_i\,
\frac{p\,m_a}{1-p(\bar f-1)}.
\end{equation}
Therefore, $M_w$ diverges when the denominator vanishes in the second term, yielding the generalized threshold condition,
\begin{equation}
p(\bar f-1)=1.
\end{equation}

\subsection{Heteroassociative system}

For a heteroassociative system composed of $n_{A,f_i}$ molecules of component A with functionality $f_i$ and $n_{B,g_j}$ molecules of component B with functionality $g_j$, we define
\begin{equation}
x_i=\frac{f_i n_{A,f_i}}{\sum_k f_k n_{A,f_k}},
\quad
y_j=\frac{g_j n_{B,g_j}}{\sum_k g_k n_{B,g_k}}.
\end{equation}
Here, $x_i$ and $y_j$ are the probabilities of reaching an A-type or B-type junction of a given functionality when following a reacted bond.

Because only A--B bonds are allowed, a reacted outward branch from an A group must terminate at a B junction. The outward expected branch weights satisfy the coupled recursions，
\begin{equation}
\begin{aligned}
E(W_A^{\mathrm{out}})&=0(1-p_A)+p_A\sum_j y_j\,E(W_{B,g_j}^{\mathrm{in}}),\\
E(W_B^{\mathrm{out}})&=0(1-p_B)+p_B\sum_i x_i\,E(W_{A,f_i}^{\mathrm{in}}),
\end{aligned}
\end{equation}
together with
\begin{equation}
\begin{aligned}
E(W_{A,f_i}^{\mathrm{in}})&=M_{A,f_i}+(f_i-1)E(W_A^{\mathrm{out}}),\\
E(W_{B,g_j}^{\mathrm{in}})&=M_{B,g_j}+(g_j-1)E(W_B^{\mathrm{out}}).
\end{aligned}
\end{equation}

Defining
\begin{equation}
\bar f_A=\sum_i x_i f_i,
\quad
\bar f_B=\sum_j y_j g_j,
\end{equation}
and
\begin{equation}
m_a=\sum_i x_i M_{A,f_i},
\quad
m_b=\sum_j y_j M_{B,g_j},
\end{equation}
the coupled equations can be solved to give
\begin{equation}
\begin{aligned}
E(W_A^{\mathrm{out}})
&=
\frac{p_A m_b+p_Ap_B(\bar f_B-1)m_a}
     {1-p_Ap_B(\bar f_A-1)(\bar f_B-1)},\\
E(W_B^{\mathrm{out}})
&=
\frac{p_B m_a+p_Ap_B(\bar f_A-1)m_b}
     {1-p_Ap_B(\bar f_A-1)(\bar f_B-1)}.
\end{aligned}
\end{equation}

Accordingly,
\begin{equation}
\begin{aligned}
E(W_{A,f_i})&=E(W_A^{\mathrm{out}})+E(W_{A,f_i}^{\mathrm{in}})=M_{A,f_i}+f_iE(W_A^{\mathrm{out}}),\\
E(W_{B,g_j})&=E(W_B^{\mathrm{out}})+E(W_{B,g_j}^{\mathrm{in}})=M_{B,g_j}+g_jE(W_B^{\mathrm{out}}).
\end{aligned}
\end{equation}

So, the weight-average molecular weight of the whole system is then
\begin{equation}
M_w=\sum_i w_{A,f_i}E(W_{A,f_i})+\sum_j w_{B,g_j}E(W_{B,g_j}),
\end{equation}
with
\begin{equation}
\begin{aligned}
w_{A,f_i}&=
\frac{M_{A,f_i}n_{A,f_i}}
{\sum_k M_{A,f_k}n_{A,f_k}+\sum_k M_{B,g_k}n_{B,g_k}},\\
w_{B,g_j}&=
\frac{M_{B,g_j}n_{B,g_j}}
{\sum_k M_{A,f_k}n_{A,f_k}+\sum_k M_{B,g_k}n_{B,g_k}}.
\end{aligned}
\end{equation}
Substituting the above expressions gives an $M_w$ with the common denominator
\begin{equation}
1-p_Ap_B(\bar f_A-1)(\bar f_B-1).
\end{equation}
Therefore, the generalized heteroassociative threshold condition is
\begin{equation}
p_Ap_B(\bar f_A-1)(\bar f_B-1)=1.
\end{equation}
These expressions reduce to the classical \fsdel{Flory--Stockmayer}\fs{F--S} results in the monodisperse limit.

\bibliographystyle{apsrev4-2-titles}
\bibliography{ref}

\newpage
\onecolumngrid
\thispagestyle{empty}

\vspace*{\fill}
\section*{For Table of Contents use only}
\begin{figure*}[hb]
\centering
\includegraphics[width=0.75\textwidth]{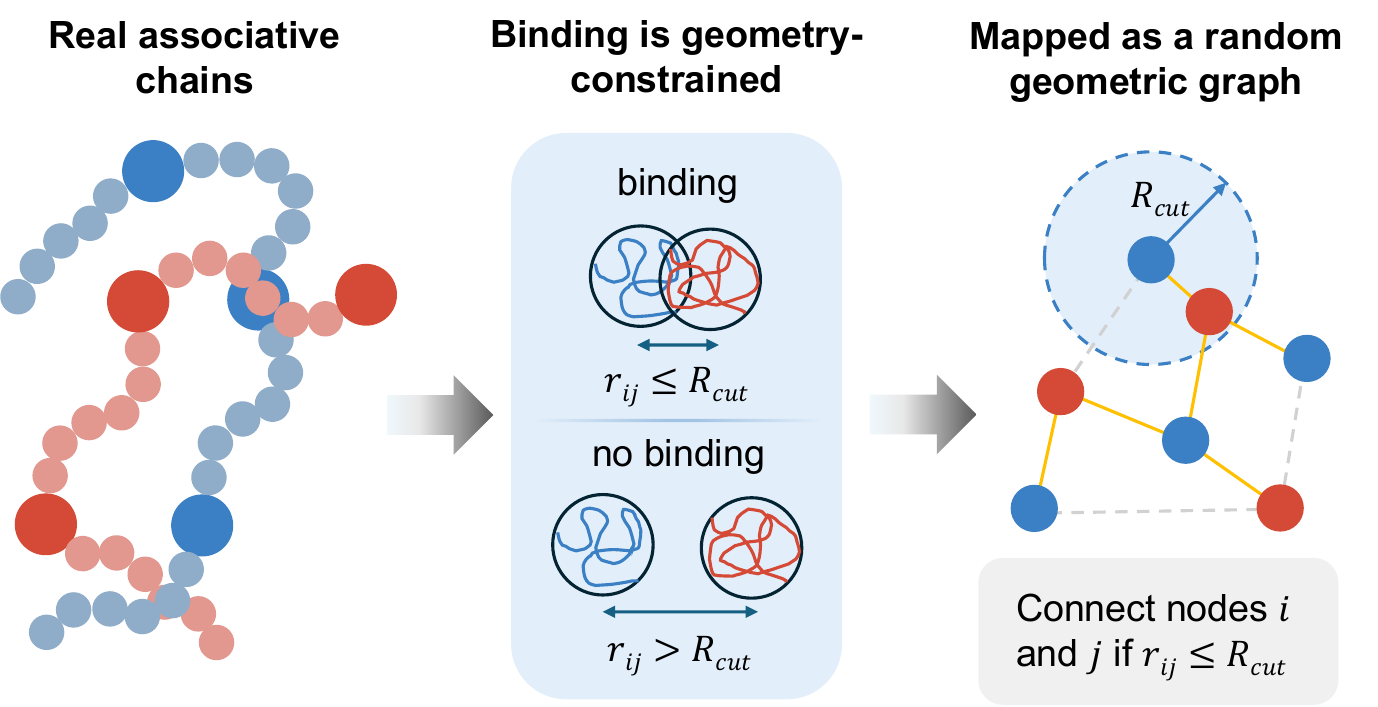}
\end{figure*}
\vspace*{\fill}

\end{document}